# Neural spikes as rare events

**Siddharth Kackar**[1,2]

1 - UCL Institute of Ophthalmology, University College London, United Kingdom

2 - Department of Physics, Blackett Laboratory, Imperial College London, London, United Kingdom

**Abstract**

We consider the information transmission problem in neurons and its possible implications for learning in neural networks. Our approach is based on recent developments in statistical physics and complexity science. Combining sensory information from various modalities for perceptual decision-making offers several advantages and is essential for the survival of both humans and animals. Not much is known about which brain regions are involved in spatial localization using audiovisual integration. We explore this further by training mice in a task requiring audiovisual integration. We then record from the secondary motor cortex (M2) using high-density electrophysiology. Analyzing this data, we found neurons responsive to multimodal as well as unimodal auditory and visual stimuli. The neurons are generally more responsive to auditory, rather than visual, stimuli. There was low correlation between the auditory and visual responses. Some neurons were sensitive to the task mode, whether active or passive, with more neurons being responsive in the active mode. A relatively large percentage of neurons (10-11%) differed significantly in their response to left and right-sided auditory stimuli, but only in 1 of the 3 mice we recorded from. These findings suggest a role for M2 in multisensory decision making and should enable further research in this field. We then use branching process simulations to model neural activity. This would support temporal coding theory as a model for neural coding.

**Introduction**

The ability to integrate information from different sources is an ethologically important process and it offers animals tremendous advantages. Spatial localization using multimodal stimuli, for example, is critical for both predator and prey in the wild and for pedestrians crossing the street.

Studies have shown that integrating information from several modalities is advantageous for perceptual decision making (Murray and Wallace, 2011). In addition, multisensory integration enhances speed and accuracy of responses (Calvert and Thesen, 2004; Stevenson et al., 2014). Accumulating evidence also shows that deficits in multisensory processing are present in psychological disorders, including children with autism (Foss-Feig et al., 2010). However, the mechanism by which the brain integrates multimodal stimuli is not yet clearly defined.

There have been several conflicting studies regarding the putative site of spatial integration, particularly with respect to audiovisual localization. In earlier studies, several brain regions such as the superior colliculus, association cortex, and the dorsal medial superior temporal area have been proposed as sites of multisensory integration (Angelaki, Gu and DeAngelis, 2009). Studies in rats have reported a role for the posterior parietal cortex (PPC) in spatial integration but inactivating the PPC did not impair behavior (Raposo, Kaufman and Churchland, 2014). The major drawback of these model organisms is that high-throughput methods are not available for them. On the other hand, the precise advantage that mice offer is that we can head-fix them and that we have a larger variety of genetic (e.g. optogenetics), psychophysical and optical imaging tools available to us. Resources such as the Allen Brain Atlas and the Mouse Brain Architecture Project make mice a particularly viable model for studying sensory processing and decision making (Carandini and Churchland, 2013). However, very few studies have examined multisensory integration in mice. One such study, by Song et al. (2017), reported an auditory dominance over visual stimuli during cross-modal activation due to feedforward inhibition of the visual cortical inputs to the PPC by the auditory cortical inputs.

Contrary to previous studies (Angelaki, Gu and DeAngelis, 2009; Raposo, Kaufman and Churchland, 2014; Song et al., 2017), we found no evidence for a multisensory role of PPC. Instead, we identified the secondary motor (M2) region of frontal cortex as being critically involved in multisensory integration as inactivating M2 impaired behavior. This project builds on the work done so far in our lab by recording from the M2 while the mice are engaged in a task requiring audio-visual integration. We used neuropixels probes to record from the mice (Jun et al., 2017). These probes offer tremendous advantages over traditional electrophysiological recording methods as the large volume coverage offered by the probes allows us to record from hundreds of sites per probe simultaneously. Additionally, the probes have a small cross-sectional area, which minimizes brain tissue damage. The dense recording sites and high channel counts give the probes a high spatiotemporal resolution, yielding well-isolated neuronal spiking activity (Jun et al., 2017).

One goal of our project was to find out how the secondary motor cortex neurons encode this integration. To do so, we trained mice to perform the task. Following the training, we recorded from the M2 using neuropixels probes. we then analyzed the behavior and spiking data on MATLAB which allowed us to build a more complete picture of the neural correlates of audiovisual integration in M2.

In building a model of the flow of information in the nervous system, we need to understand the relationship between a stimulus and the response it evokes in a single neuron or network of neurons. In addition, we also need to understand how that stimulus is represented or coded in the neuron and how that representation relates to observed behavior. Neural coding is concerned with studying these two aspects of information flow. We want to use our current knowledge of neural coding and information theory to study how the brain perceives the statistical features of the input and encodes them to form representations of our world. Also, we want to study how the brain stores this information in the longer term i.e., how it learns from probabilistic stimuli. Ultimately, we want to contribute toward a unified understanding of perception and learning. This will require the combined efforts of electrophysiologists, molecular biologists, and cognitive scientists and we want to contribute to this problem from a computational/theoretical standpoint.

While much progress has been made in mapping neural correlates of sensory or motor variables, we still need to understand how the microarchitecture of brain circuitry supports cognitive processes underlying thinking, memory, and decision-making. A particularly important challenge is to clarify the operating principles governing how information processing is coordinated over time across large spatial scales. The human brain is far “smarter” than a current generation supercomputer yet consumes 100,000-fold less space and energy (Sterling & Laughlin, 2015). Neural dynamics have been observed to show a temporal mode of emergence. Long periods of quiescence are followed by sudden bursts of activity (Tagliazucchi et al., 2012). This pattern of activity may be associated with consciousness (Paradisi et al., 2013). Principles from statistical physics and information theory may provide powerful tools for revealing these operating principles.

In recent years, the statistical physics of avalanches – as observed in earthquakes, sandpiles, and forest fires – has been used to explain the dynamics of bursts of neuronal activity observed spontaneously in brain tissue in culture and in vivo (Beggs & Plenz, 2003; Hahn et al., 2010; Petermann et al., 2009). It is well known that brain dynamics at the neuronal level usually involve intermittent bursts of activity in the form of avalanche-like phenomena spanning a wide range of spatiotemporal scales. However, there is an important knowledge gap in establishing how this intermittent phenomenon underpins higher brain function. Aligning with the theory of critical branching processes, the propagation of these bursts follows a power law with an exponent of -3/2 for event sizes, with a branching parameter close to the critical value of 1 (Christensen & Moloney, 2005; Jensen, 1998, 2022). This has been suggested to optimize information transmission in networks, while preventing runaway network excitation, with such avalanches providing a signature of a multiscale self-organizing process. The precise operating point, near to criticality, used by neural circuits remains a matter of debate. Slice physiology indicates the balance of excitation and inhibition found in vivo seems to favor self-organized criticality (Shew et al., 2011). It has also been conjectured that it may be advantageous for brain networks to operate just below criticality (Beggs, 2022; Williams-García et al., 2014).

However, a crucial unanswered question is how neuronal avalanches affect behavior, particularly cognitive phenomena such as memory and decision-making which we might expect to be driven by internal processes. Altered avalanche dynamics have been implicated in cognitive dysfunction caused by schizophrenia (Seshadri et al., 2018). The emergence of technology for simultaneously monitoring up to tens of thousands of neurons across the brain (Demas et al., 2021; Stringer et al., 2021) makes it timely, and crucially important, to develop theoretical approaches to make sense of the brain-scale coordination of information processing critical to understanding cognitive function.

A vital model studied in complexity science is that of a branching process, also known as an avalanche or chain reaction. Consider the spreading of infectious diseases. It has been shown that if the average number of people infected by each patient approaches 1, the total size of the chain of infected agents follows a power law with an exponent of 3/2 (Jensen, 1998, 2022). Many examples of this exponent have been observed in nature, such as the spreading of brain activity both at the level of individual neurons (Beggs & Plenz, 2003) and neural populations (Tagliazucchi et al., 2012) and forest fires (Song et al., 2006).

In a recent paper, Fosque et al. tried to use the concept of criticality to look for biomarkers of brain health (Fosque et al., 2022). They have analyzed data from magnetoencephalography (MEG) studies of the human brain. They have used the quasi-criticality framework (Williams-García et al., 2014), which states that the effective critical exponents for brain function can be considered to lie along a scaling line. They hypothesize that the position of the exponents and its change along this scaling line can be used as a biomarker for neurological aging and health.

It is our opinion that this correlation between aging and the change in the position of the biomarkers along the scaling line cannot be entirely correct. Essentially, they are using the failure of the scaling relation as a measure of distance from criticality. However, it is known that the exponents of a scaling relation do not change. Their results may be an artifact due to a lack of precision in their experiments.

## Information theory and neural coding

Information theory provides mathematically rigorous tools to quantify the precision of information transmission, setting theoretical limits on maximum information capacity (Borst & Theunissen, 1999). The three central questions in the field of neural coding are to find out:

A. What is being encoded? For example, whether the information is encoded in the amplitude of spikes or the change in amplitude?
B. How is it being encoded? The question of rate vs temporal coding has seen conflicting reports, with conventional studies supporting rate coding as standard (Adrian & Zotterman, 1926), but more recent studies report examples of temporal coding (Stein et al., 2005).
C. With what precision is it being encoded? What is the degree of variability in the responses?

Neuroscientists have traditionally addressed the first two questions by studying stimulus-response curves and changing the stimulus ensembles and response measures. Error bars were used to assess the variability. Information theory, on the other hand, provides mathematically rigorous tools to quantify the precision of information transmission, setting theoretical limits on maximum information capacity. The approach can be applied to study neural coding in the following ways:

A. Estimating the maximum information capacity (channel capacity) of a neuron and the actual information transmitted, to quantify efficiency.
B. Compare the upper bound of information transmitted with that of an optimal linear model, to test for non-linearities and find out how a neuron transforms our data.
C. Determining the limiting temporal precision of code, i.e., the minimal timescale in which information is contained, to find out if there is any information in the precise timings of the spikes.

To calculate maximal information transfer, a new quantity was introduced, called entropy. Entropy characterizes how many free states a variable can assume and the probability of each i.e., the variability. Entropy is the information needed to eliminate all uncertainty about a variable. Information can be considered as a reduction in entropy. Three common approaches are used for estimating the mutual information between the stimulus and response, the direct method, which calculates the exact value of information, and the upper and lower bound methods. Factors such as the experimental parameters and quality of stimulus ensembles determine our choice of method. Generally, a combination of all three is used as an optimal linear model.

Another area of application of information theory is population coding. Population coding is defined as a method to represent stimuli from multiple neurons. The response of each neuron has a probabilistic distribution over a set of stimuli, which is considered together with other neurons to characterize certain features of the input. Population analysis is reported to have several advantages over single-neuron recordings in reducing uncertainty due to neuronal variability and the ability to represent different attributes of the stimulus simultaneously.

Studies have suggested using linear decoding algorithms for population recordings in conjunction with information-theoretic tools as mutual information gives a more comprehensive quantification of the information contained in a neuronal population, by evaluating the reduction of uncertainty about the stimuli that can be obtained from the neuronal responses (Quian Quiroga & Panzeri, 2009). Representing uncertainty is important for decisions involving risks and may be fundamental for neuronal computations in the presence of sensory and neural noise.

**The information bottleneck method**

The information bottleneck method  (Tishby et al., 2000) tries to answer how much relevant information survives a communication. Suppose there is an input variable X with a hidden characteristic Y. To transfer relevant information about Y from X to the output state T, we try to maximize the mutual information between Y and T, which is bounded above by the mutual information between X and T (by the data processing theorem), which we are trying to minimize. What this means is that we are trying to preserve the maximum amount of relevant information about Y while letting go of the unnecessary parts of X. This creates the "information bottleneck" in the T-X-Y Markov chain.

Experiments on the H1 motion-sensitive neuron of the fly visual system tried to characterize the neural "dictionary" containing the relevant stimuli and their respective responses, using the information bottleneck (Schneidman et al., 2001). They tried to quantify the compressibility of said dictionary i.e., they tried to find out which features of the stimuli are relevant. They found that the stimulus code was highly compressible meaning the neuron was sensitive to only a few significant features. This feature of compressibility was preserved in different flies.

The information bottleneck method also has applications in understanding deep learning. Each layer of a deep neural network can be treated as input and output points for the surrounding layers, requiring analysis of information compression by the bottleneck method (Schwartz-Ziv & Tishby, 2017; Tishby & Zaslavsky, 2015). However, recent studies have argued against this method being used to explain deep learning saying that the results of Schwartz-Ziv and Tishby are particular to their setup and do not hold in the general case (Saxe et al., 2019). Thus, this area is of intense interest in neuroscience and machine learning communities.

**Information bottleneck and the renormalization group**

In the context of neuroscience, both the information bottleneck and the renormalization group are coarse-graining methods used to analyze the flow of information between the layers of a neural network (Meshulam et al., 2019). In another paper, they presented the renormalization group and the information bottleneck in a unified framework (Tan et al., 2019). Their renormalization procedure achieves a coarse-graining where they can choose which relevant features to keep, for example, by keeping information about long-range interactions while removing local information. Other studies have drawn parallels between the coarse-graining performed by neural networks and renormalization group procedures (Iso et al., 2018), found similarities in the notion of “relevance” in renormalization and the information bottleneck (Gordon et al., 2021), and have devised a self-supervised learning method based on combining the renormalization group and the information bottleneck (Ngampruetikorn et al., 2020).

## Methods

**Surgery and training:** To understand the role of M2, we first trained mice in a task requiring audiovisual integration. All the mice used were female. We used 5 mice for the actual recording which required training 6 mice in a modified version of the 2- alternative forced-choice task described by Burgess *et al.* (2017). Prior to training, the mice underwent surgery. Isoflurane (Merial; 3.5% for induction and 1-2% during surgery) and lidocaine (local; 6mg/kg, Hameln pharmaceuticals ltd) were used for anesthesia. Carprofen (5 mg/kg; Rimadyl, Pfizer) was administered intraoperatively via the subcutaneous route for systemic analgesia. Eyes were covered with chloramphenicol (Martindale Pharmaceuticals Ltd). The animal was placed into a stereotaxic apparatus and the skin and connective tissue surrounding the area of interest were removed. A custom-made head-plate was positioned above the area of interest and attached to the bone with Superbond C&B (Sun Medical). Throughout all surgical procedures, the body temperature was stabilized at 37°C by heating pads. Subcutaneous injections of 0.01 ml/g/h of Hartmann's solution were given. After the surgery, the animal was placed into a heated cage for recovery from anesthesia. Mice were given three days to recover while being treated with carprofen.

The mice were water-restricted to encourage adaptation to the training rig and increase motivation for water reward. Mice were mostly trained for 5 days a week. Both the training and experimental rigs had three screens and seven speakers. On each trial, the mice were head-fixed and were presented with either a visual or an auditory stimulus (unimodal) or both simultaneously (cross-modal). The stimuli could be either from the same (coherent trials) or opposite (conflict trials) directions. The mice were required to turn the wheel in order to center the stimuli. The visual stimulus was a gabor with different contrasts (10% to 80%). The auditory stimulus was filtered pink noise at 8-16 kHz. The stimuli were flickering at 8 Hz. There was an inter-trial interval of 1-2 seconds. The performance of the mice on the task has been well established by previous work done in our lab. The task was well suited for this purpose as it is closed-loop, allowing rapid learning; and is compatible with neuropixels recordings. The behavioral performance was examined using classic psychophysical analyses, including psychometric curves.

**Recordings:** Once the mice were able to perform the task to a reasonable criterion, which typically took around 4-6 weeks of training, they were shifted to the experimental rig. After an adaptation period of about 10 days, recordings were performed from the M2. On the day of the recordings, mice were briefly anesthetized with isoflurane while two craniotomies were made above the areas of interest (M2), 2 mm anterior and 1.5 mm lateral to the bregma., either with a dental drill or a biopsy punch. After at least three hours of recovery, mice were head- fixed in the setup. Probes had a soldered connection to short external reference to ground; the ground connection at the headstage was subsequently connected to an Ag/AgCl wire positioned on the skull. The craniotomies, as well as the wire, were covered with a saline bath. Probes were advanced through the saline and through the dura at a 45$^{o}$ angle to the horizontal., then lowered to their final position at ~10µm/sec. Electrodes were allowed to settle for ~5 min before starting recording. Recordings were made in external reference mode with LFP gain = 250 and AP gain = 500. We recorded both during the task and during the passive presentation of stimuli, to see if neural responses are different in M2 while the animal is engaged in the task. The passive mode had the same stimulus presentations as the active mode of the task, but the water reward was switched off.

These craniotomies allowed us 5-7 days of recordings from the site, before the growth of bony and scar tissue obstructed the site completely. We then performed a second set of craniotomies near the first ones, to record for a further 5-7 days. Following this second set of recordings, the mice were culled.

**Data analysis:** Processing the behavioral and electrophysiological data involved manually sorting the good spikes from bad ones, using Kilosort (Pachitariu *et al.*, 2016). The automated processing already labelled the clusters (spikes from a single neuron) which it thought to be noise. The manual stage involved checking these decisions and classifying the other clusters into noise, multi-unit activity (MUA) or good spikes. The decision process involved 3 stages – checking the amplitude (low amplitude not good), the shape of the waveform (noise-like?) and the presence of refractory period on the correlogram (MUA/noise had no or very small refractory periods). Generally, clusters meeting at least 2 of the 3 criteria were labelled good. Only these good spikes were used for my study. These were processed further, and the resulting data was used for analysis using MATLAB. Overall, 8166 neurons were included in the passive data and 8043 were included in the active data (neuron clusters which do not spike at all were excluded automatically by the processing software). Not all these clusters were from the M2.

As a first step in analysis, we looked at the fraction of neurons responding to auditory and visual stimuli respectively. This analysis was done in passive recordings as the active mode of the task had an onset tone from the middle speaker in all trials, which would made it difficult to differentiate the effect of visual stimuli. To compare the change in firing rate in response to auditory stimuli, a time window of 100-500 milliseconds before stimulus onset and 50-100 ms after the stimulus onset was used. For visual stimuli, the window after the stimulus onset was changed to 100-150 ms, as visual responses were observed to be slower as compared to auditory responses. A paired t-test was used to detect the changes in response for every trial, with a significance level of $p < 0.01$.

We then looked for any correlation between auditory and visual responsive neurons. This analysis was also done in passive mode with the same time windows as above. Only the neurons which were marked as responsive to either auditory or visual stimuli, from the above analysis, were included here. A best-fit line was plotted and $R^2$ and p values were calculated.

Finally, we looked at whether there were neurons whose responses differed significantly when the stimuli were presented from either the left or the right. This analysis was done for both active and passive recordings. Again, only the neurons responding to either auditory or visual stimuli were included. For the active mode, trials in which the mouse did not respond (timed out) were excluded. An unpaired t-test was used, with a significance level of $p < 0.01$.

**Simulations:** We first simulated neural activity as a branching process using a random number generator as our starting point, in Python. Analyzing burst activity is a prototypical problem in data science. We consider the neural activity as a memoryless Markov process, with each new state being independent of the previous ones. The simulations are similar to Markov Chain Monte Carlo simulations. We then measured the critical exponents and rationalized the data with respect to known properties of scaling relations.

## Results

We trained the mice in a 2-alternative forced choice (2-AFC) task, in which they moved a wheel in order to center the auditory and/or visual stimuli. The total training period for each subject was around 4-6 weeks. This time has been found in our lab to be optimal as the animals are well trained, and it avoids the pitfalls of a shorter training period. These include highly variable behavioral performance, which makes it difficult to extract any relevant information from the data. Also, well-trained animals have little confusion about the task structure, and most of the errors they make stem from confusion in their perception of the stimulus. This makes the behavioral performance easier to analyze, leading to the well-defined psychometric curves as described below (Fig. 1 (d)).

Fig. 1 (c) shows the learning rates of all the mice at the various stages of training. I began training the mice with multisensory stimuli from the same direction i.e., coherent (no conflict) trials. When the mice performed well, unisensory auditory and visual stimuli were introduced successively. These were later followed by blank trials and conflict trials, for which there was no correct response. The mice learn quickly and generalize to visual and auditory trials. Out of the 6 mice I trained, I failed to learn the task to a satisfactory standard and was removed from the study. Fig. 1 (d) shows the fraction of rightward choices made by the mice in various trial conditions. The three lines showing the performance in the auditory left, right, and center are clearly separated, showing that mice combine the information from the auditory and visual stimuli while making decisions. The grey line, showing the auditory amplitude in the middle, is centered at the origin, indicating minimal bias.

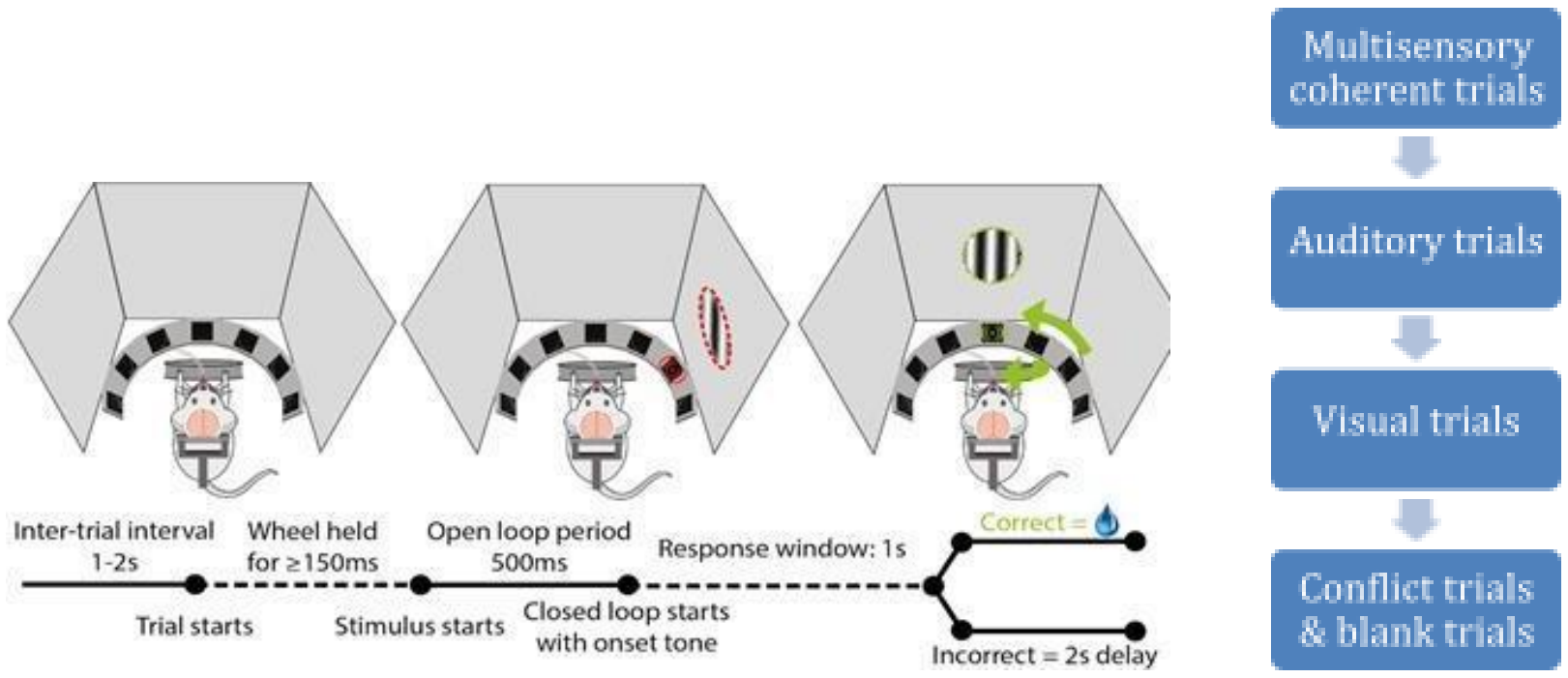

(a) Experimental set up & timeline

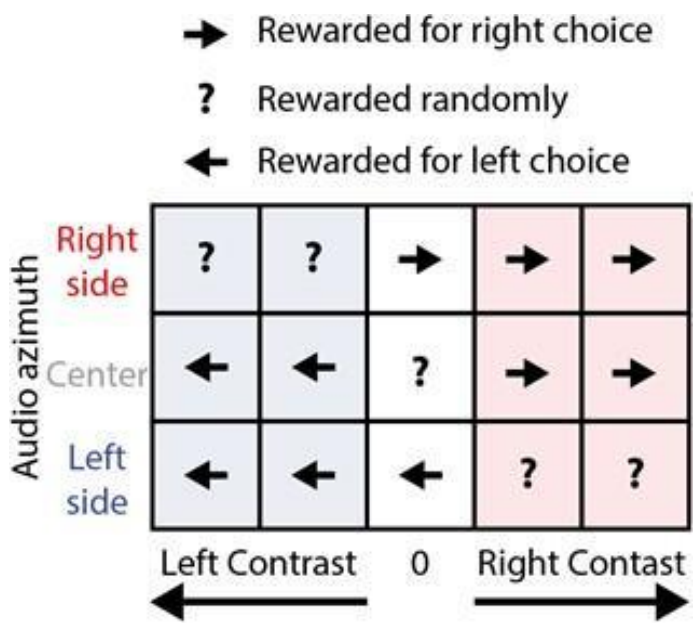

(b) Reward structure

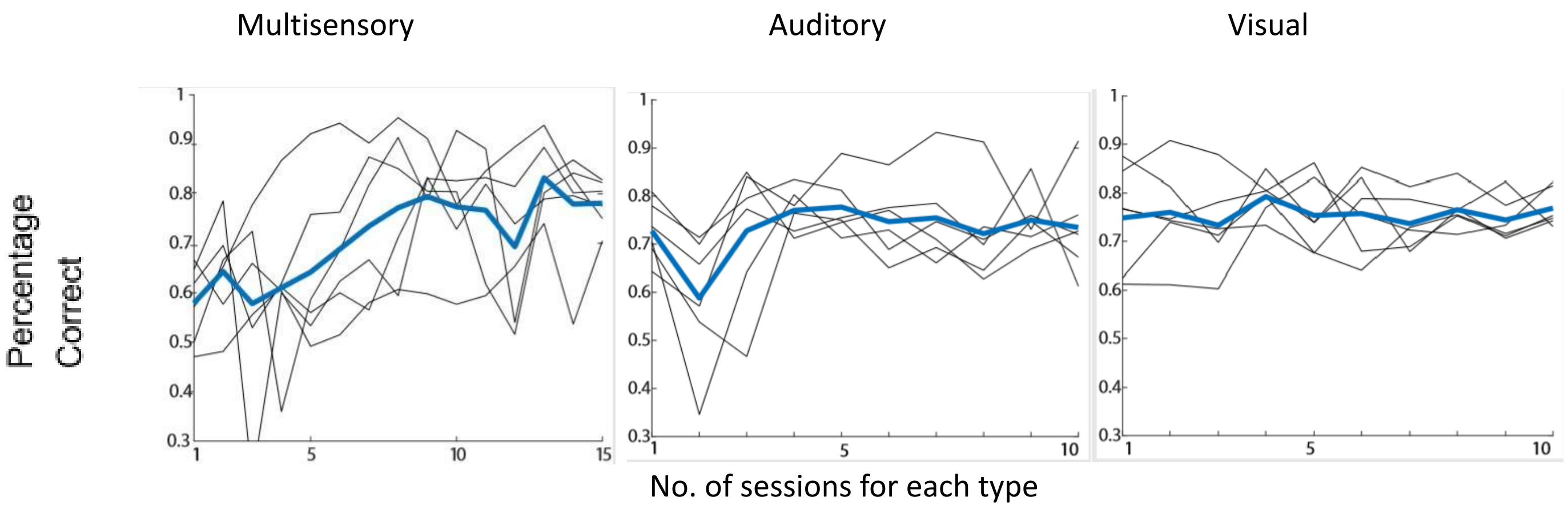

(c) Learning rate

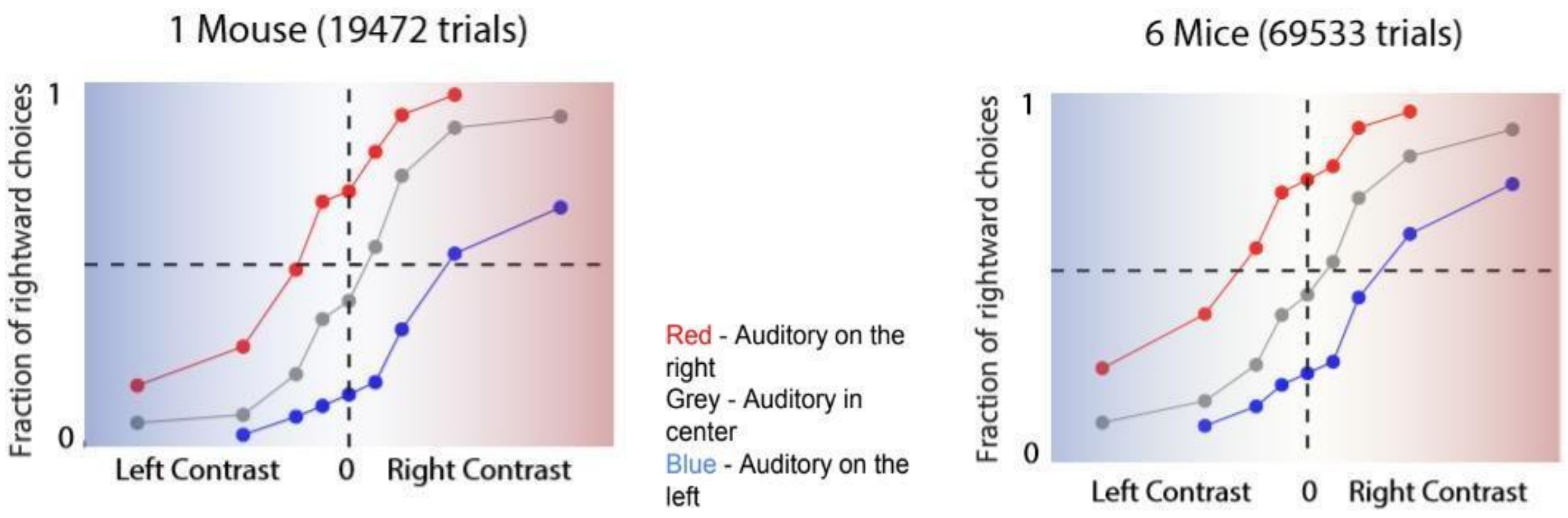

(d) Training performance

**Fig 1. Training mice in the task (a)** The experimental set-up and the timeline for the task. **(b)** The reward structure: the X-axis is the visual contrast in either direction and the Y-axis is the auditory stimuli. The mice get rewarded for correctly centering the unidirectional stimuli. For bidirectional/conflict stimuli as well as blank trials (which have an onset sound from the middle speaker) the mice get rewarded randomly. At the end of the training period, all mice were able to perform the task to a reasonable criterion ($\geq$ 75% correct responses). **(c)** The training performance of the mice at various stages of learning: we started with multisensory coherent trials, followed by auditory and then visual trials. The grey lines represent the individual mice, while the blue line is the average performance of all mice for each trial type. **(d)** The training performance of the mice expressed as fraction of rightward choices: the blue background represents increasing visual contrast for visual stimuli on the left, while the red background represents the same on the right. Blue lines on red background, and vice-versa, represent conflict trials, while lines on the same colored background are coherent trials. The example mouse performed 19472 trials, while overall, the mice performed 69533 trials.

The visual stimuli in our task had varying contrasts, from 10% to 80%. The auditory stimuli also varied in the direction of presentation. This allowed us to differentiate the effects of any bias or guessing by the subjects while performing the behavior. As a result, we obtained psychometric curves that spanned most of the Y-axis, showing good training and high engagement of the subjects. The grey line was also centered at the origin, showing a minimal bias. These are indicative of a well-performed psychophysical experiment (Carandini and Churchland, 2013).

We used neuropixels probes to record from ~8000 neurons from 3 mice (we had to cull two trained mice early as they developed ocular infections). Inserting two electrodes at a time yielded simultaneous recordings from hundreds of neurons in each recording session. In total, we performed 43 insertions in the 3 mice. We identified the firing times of individual neurons using Kilosort (Pachitariu *et al.*, 2016) and determined their anatomical locations by combining electrophysiological features with the histological reconstruction of fluorescently labeled probe tracks.

We wanted to find out if M2 neuronal activity can predict performance better than just stimulus presentations. The first step was to check whether the neurons respond to auditory or visual stimuli and with what frequency. Out of 8166 neurons, which were included in passive, 2195 responded to either auditory or visual stimuli. Fig. 2(a) shows that neurons were consistently more responsive to auditory stimuli for all 3 mice. Then, we combined the spiking data with the data from the behavioral task. We analyzed this data using MATLAB to see if there is any correlation between the behavioral decisions of the mice and the activity of the M2 neurons while performing the audio-visual integration task. Fig. 2(b) is a correlation plot between the average change in firing rate for every neuron cluster for auditory and visual stimuli. We did not observe a high correlation ($R^2$ = 0.0702) even though the p-value was very highly significant.

In the next two figures (Fig. 2 (c) and 2 (d)), we tried to find out if there were neurons that responded significantly differently to stimuli from the left and right directions, in both active and passive modes. In active mode, 3161 out of 8043 total neurons were responsive to either auditory or visual stimuli. What I found was that the percentage of such cells was generally low for the mice, except for mouse 2, which had 10% and 11% of neurons responding differently to auditory stimuli from the two sides in active and passive recordings, respectively (7% and 2.8% for visual stimuli). For an example of such a neuron, see the multimodal neuron in Fig. 2 (c), which responds differently for auditory stimuli from both sides.

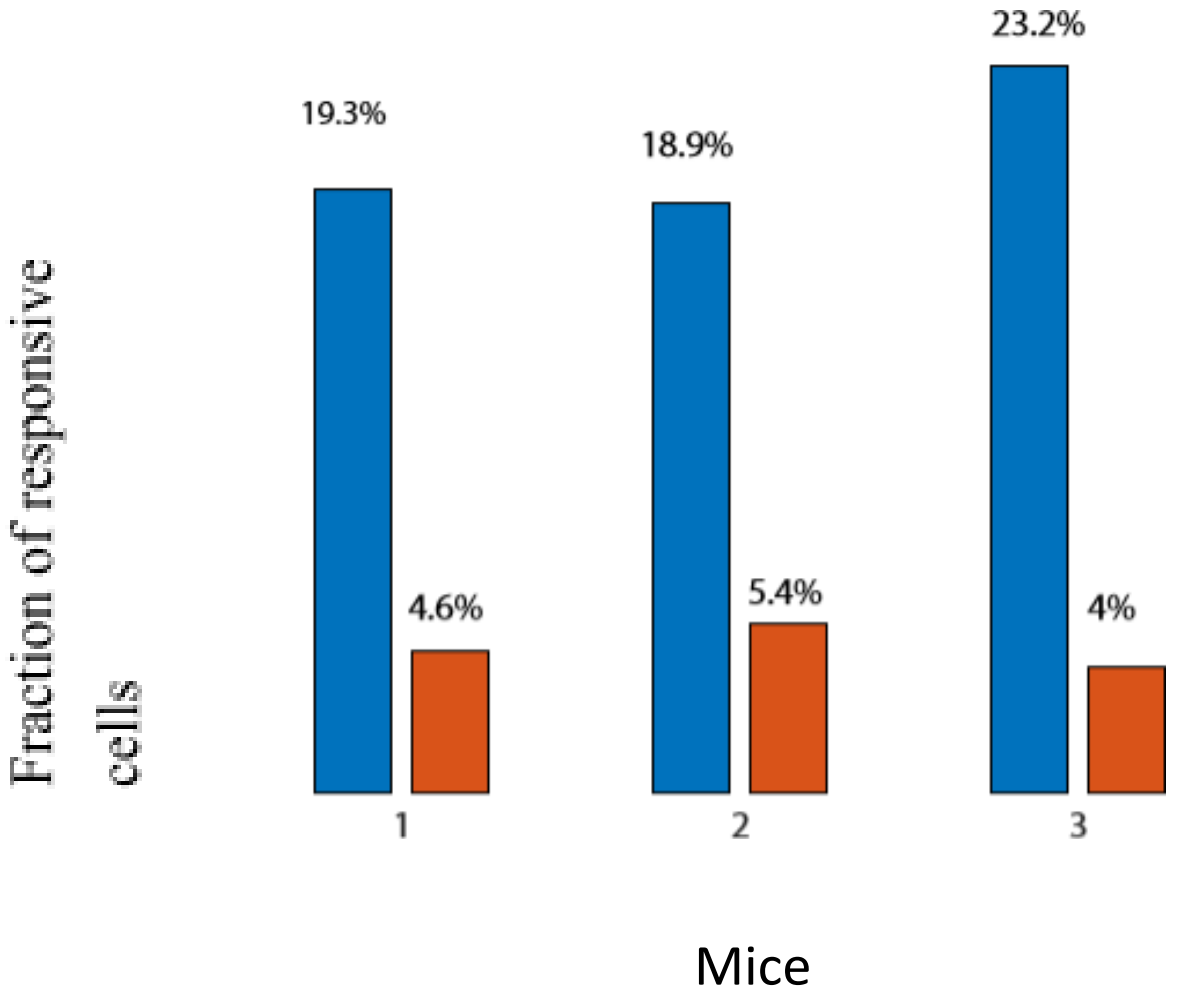

(a) Fraction of cells responsive to auditory (blue) and visual (orange) stimuli in passive

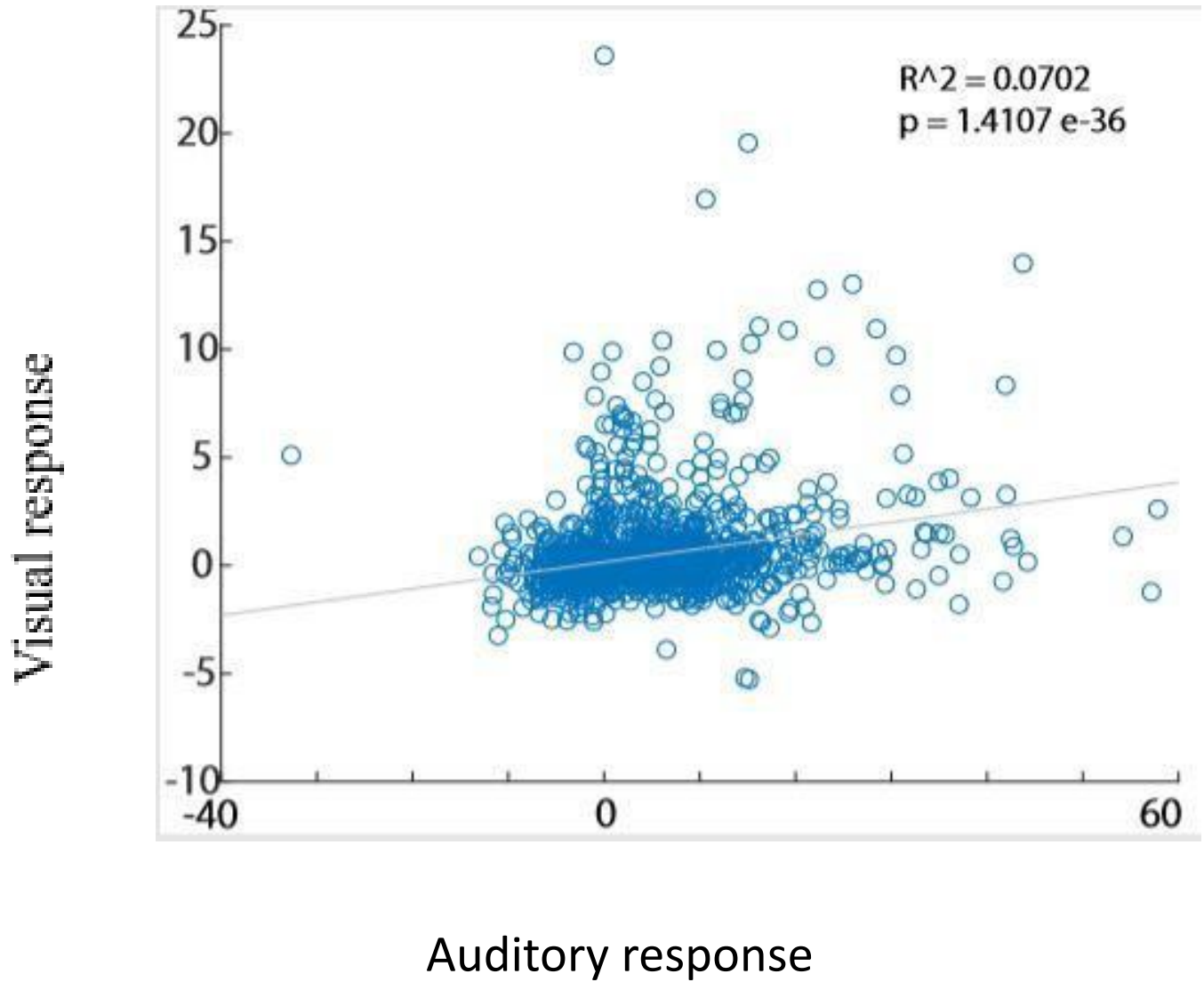

(b) Correlation plot between response to visual and auditory stimuli in passive

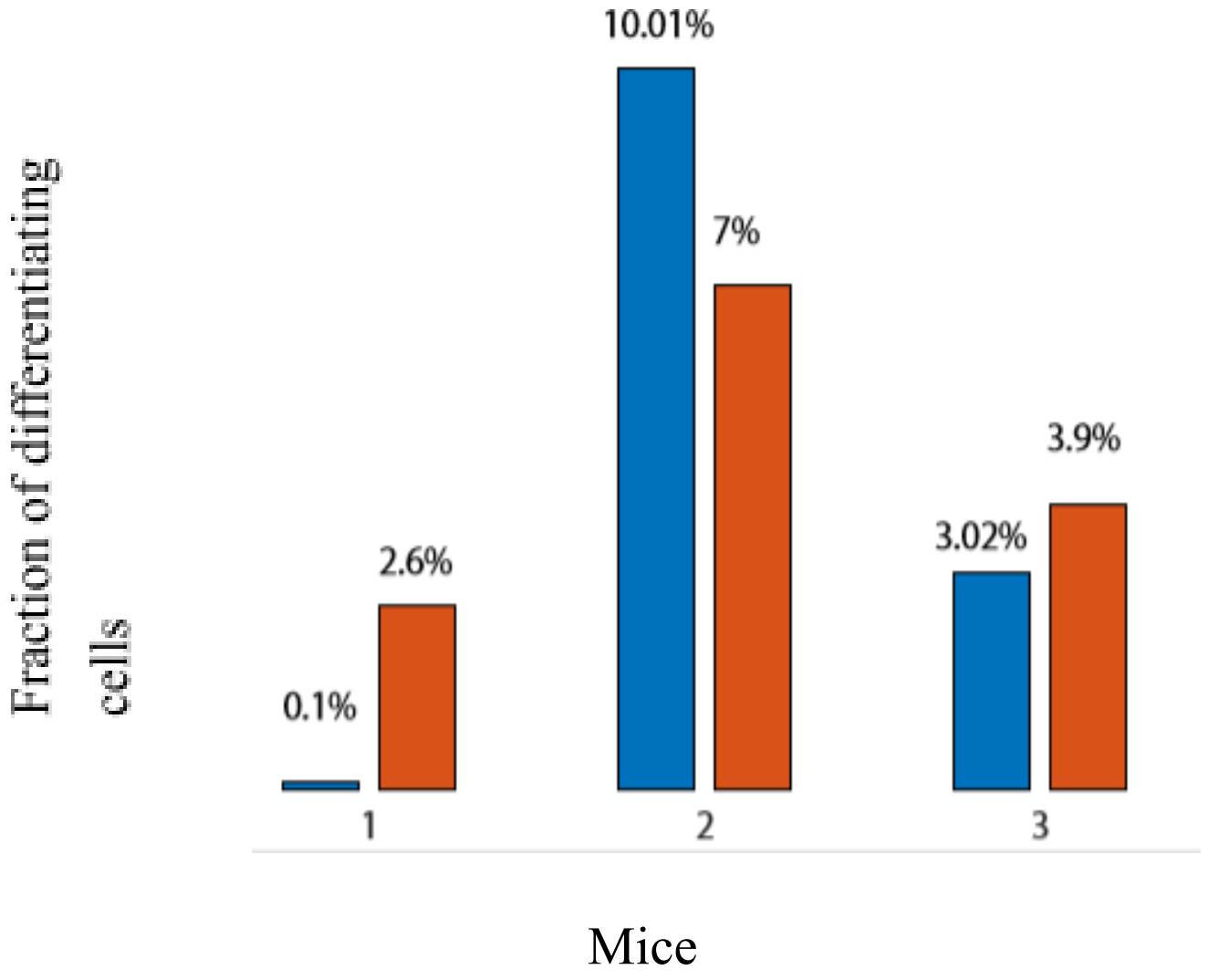

(c) Fraction of cells responding significantly differently to left and right auditory (blue) and visual (orange) stimuli in passive recordings

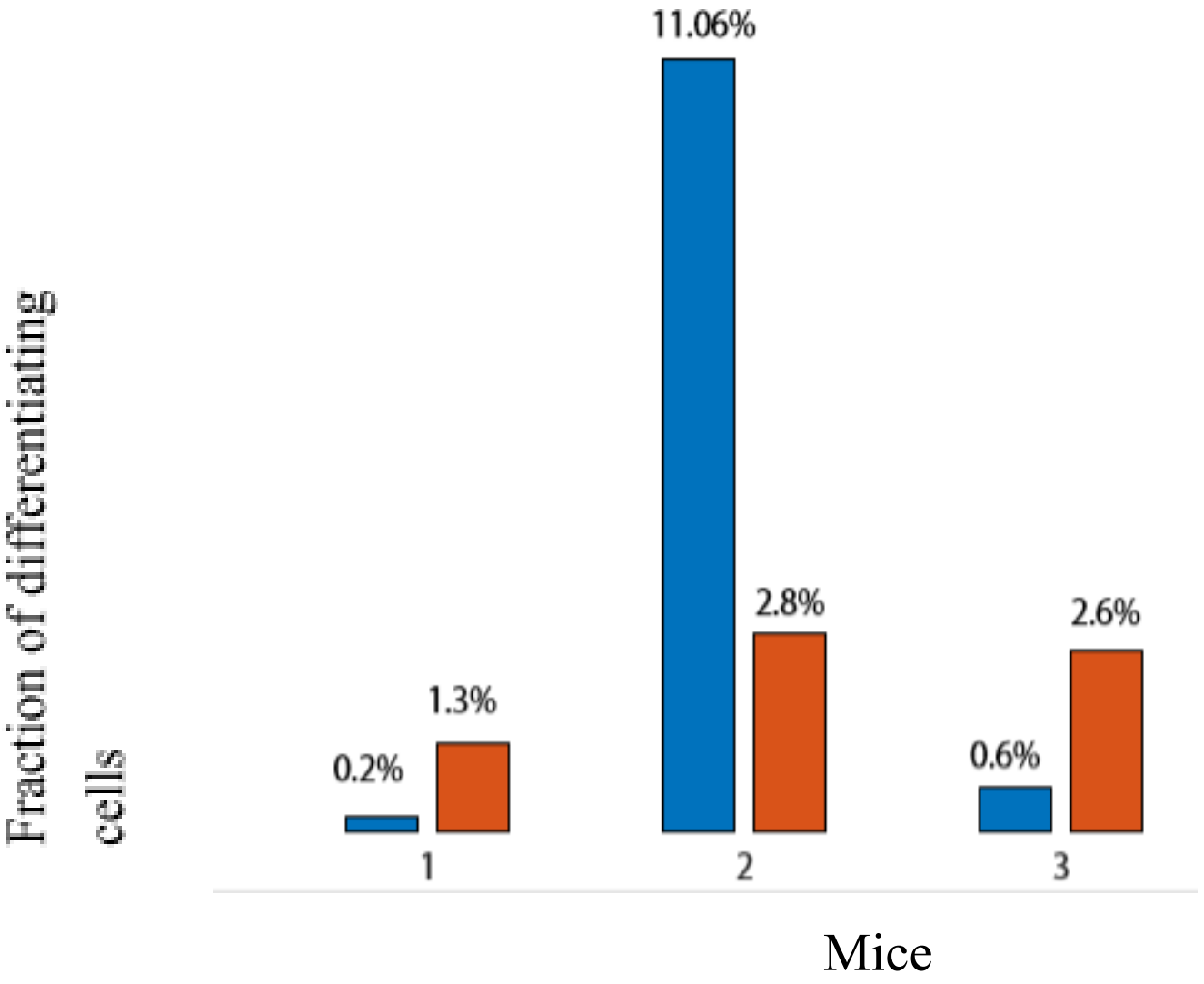

(d) Fraction of cells responding significantly differently to left and right auditory (blue) and visual (orange) stimuli in active recordings

**Fig. 2: Analysing the data (a)** Fraction of cells responding to auditory and visual stimuli. The Y-axis is the fraction of responsive cells and on the X-axis are the individual mice, with the neuron clusters divided into auditory (blue) and visual (orange) responsive cells. Paired t-test was used for analysis, with significance level of $p<0.01$. 8166 cells were included in the passive data, out of which 2195 were responsive to either stimulus. **(b)** Correlation plot between response to visual and auditory stimuli. The Y-axis represents the visual responses and the X-axis is the auditory responses. Only the 2195 responsive neurons from the earlier analysis were included. $R^2$ was 0.0702, indicating low correlation between auditory and visual responsive neurons. **(c) & (d)** Fraction of cells which respond differently to stimuli from the left and right directions. In active, the 3161 responsive neurons were included. Unpaired t-test with significance level of $p<0.01$ was used. (n = 3 mice)

A vital model studied in complexity science is that of a branching process, also known as an avalanche or chain reaction. A scaling relation is an equation used to describe a branching process. It is known that the exponents of a scaling relation do not change. In order to consolidate our view, we have first simulated neural activity as a branching process using a random number generator as our starting point. Analyzing burst activity is a prototypical problem in data science. We will consider the neural activity as a memoryless Markov process, with each new state being independent of the previous ones. The simulations are similar to Markov Chain Monte Carlo simulations. We have measured the critical exponents and rationalized the data with respect to known properties of scaling relations.

Here S denotes the size of an avalanche, T denotes the time duration, and log<S>(T) denotes log plots of all avalanches for a particular time duration T. Tau S, tau T, and gamma are the three critical exponents. The critical exponents that we have obtained from our branching process simulations are approximately twice the expected value from analytical studies. Possible reasons for this could be that currently, it is not very clear as to which generator function to use for such studies, Poisson or geometric. Here, we have used a Poisson distribution as the results were closer to the analytical results.

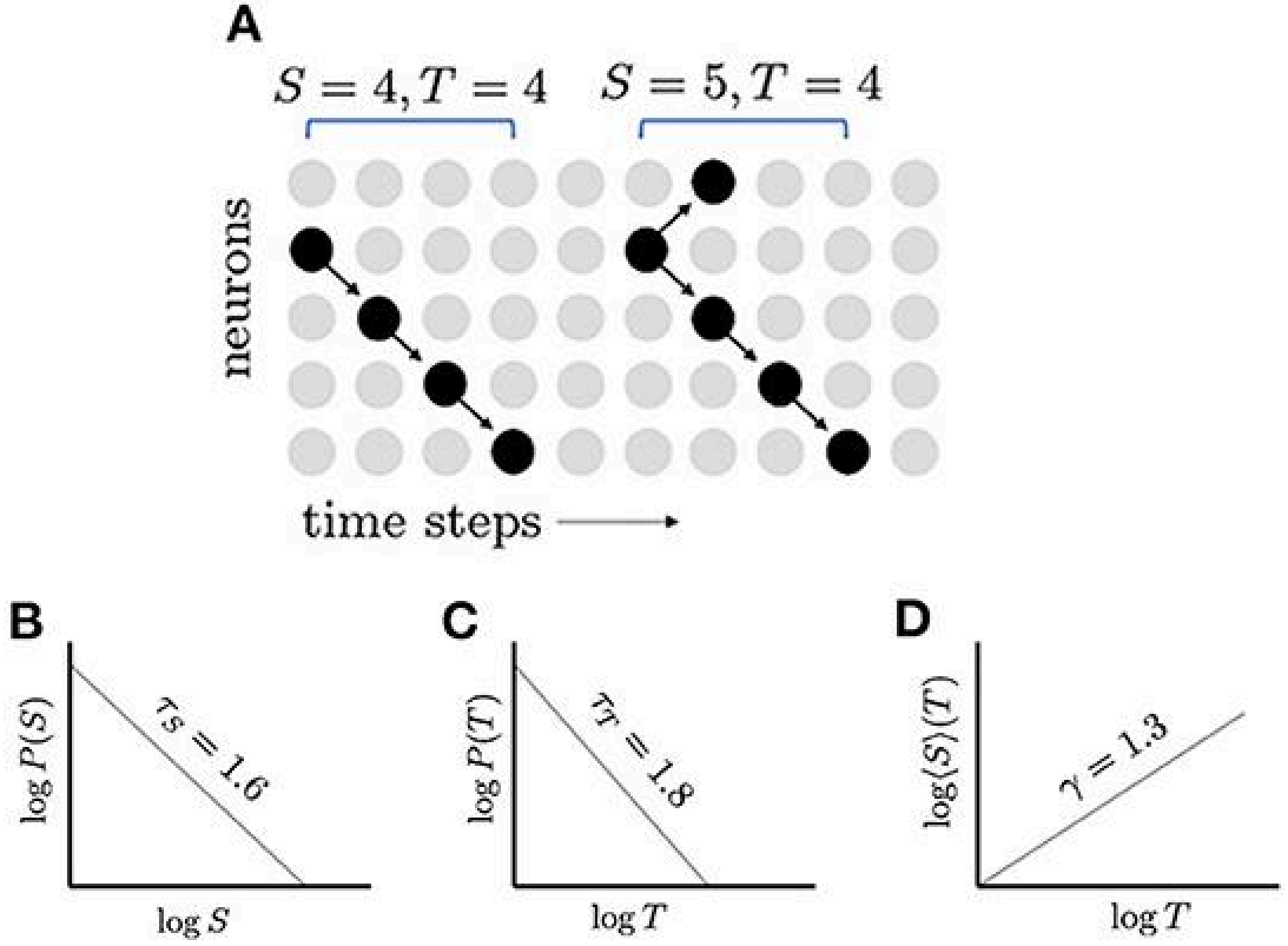

**Fig. 3:** Known critical exponents (Fosque et al., 2022)

**Table 1:** Branching Process (lambda from 0.85 to 1.05)

| **Lambda** | **Tau_S** | **Tau_T** | **Gamma** | **Predicted Gamma** |
|---|---|---|---|---|
| 0.85 | -3.743 | -3.69 | 1.449 | 0.988 |
| 0.9 | -3.714 | -3.554 | 1.513 | 0.965 |
| 0.95 | -3.701 | -3.812 | 1.508 | 1.023 |
| 1.0 | -3.698 | -3.71 | 1.761 | 1.002 |
| 1.05 | -1.345 | -3.465 | 1.531 | 1.904 |

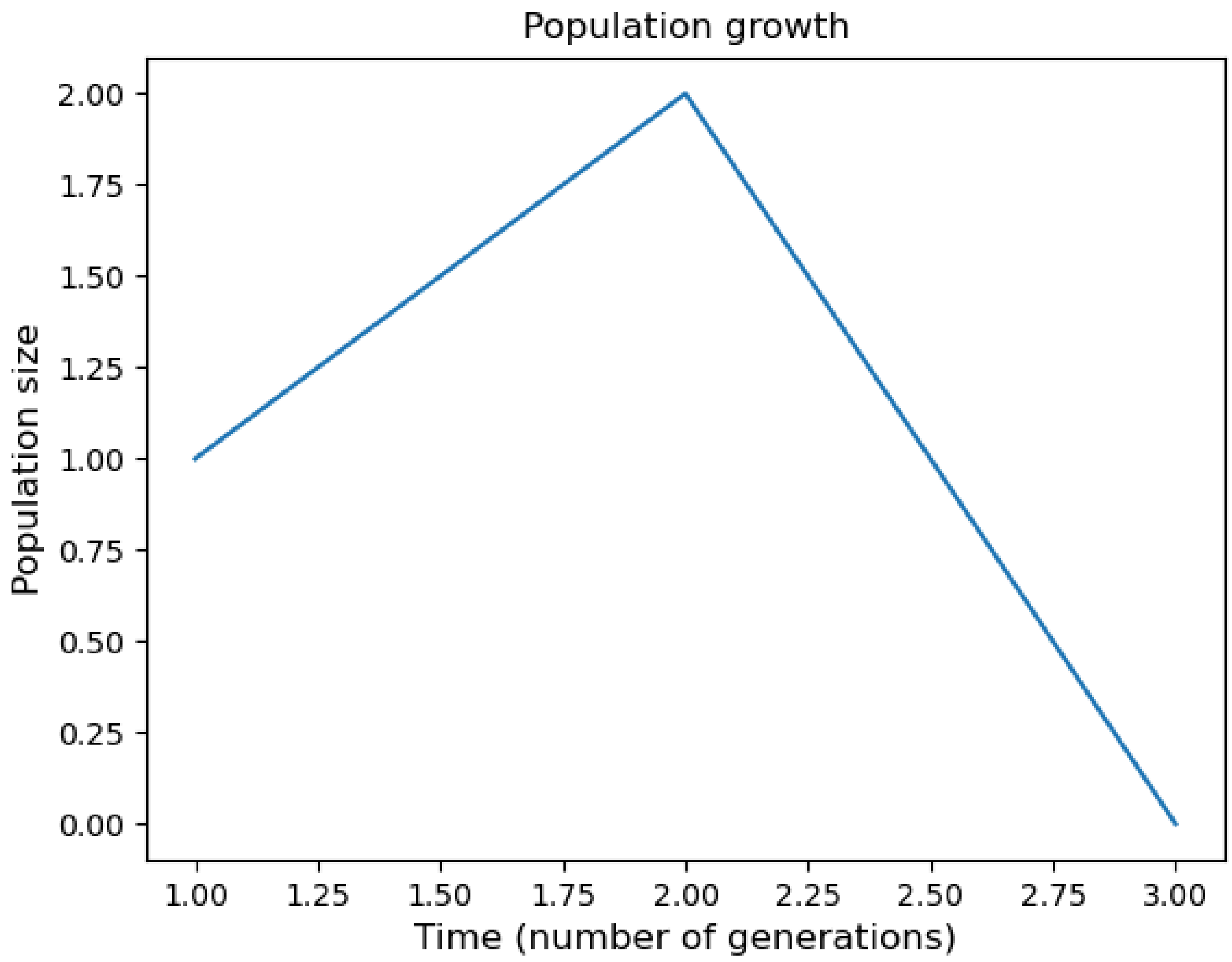

**(a)**

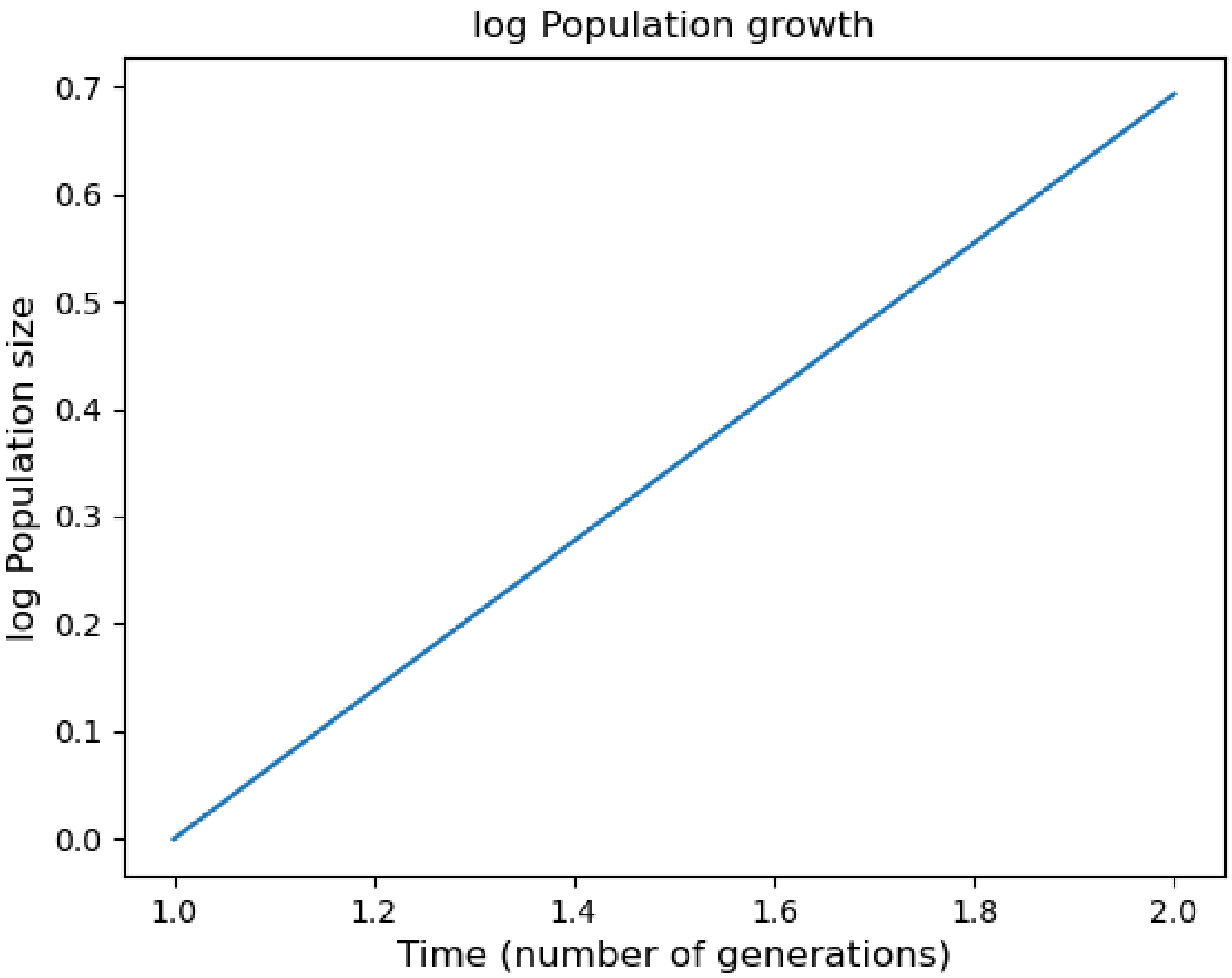
log Population growth
log Population size
0.0
0.1
0.2
0.3
0.4
0.5
0.6
0.7
1.0
1.2
1.4
1.6
1.8
2.0
Time (number of generations)

**(b)**

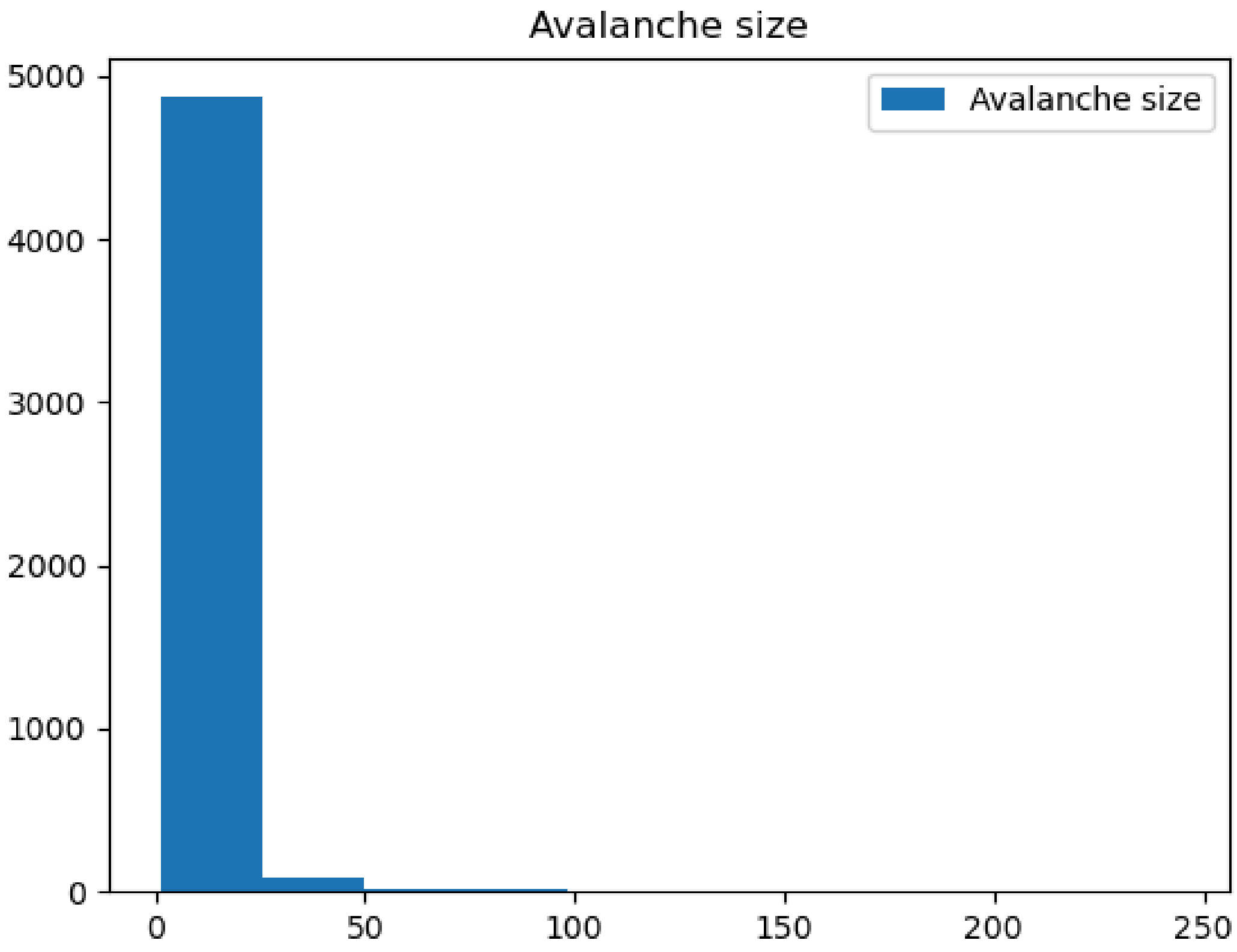
Avalanche size
Avalanche size
5000
4000
3000
2000
1000
0
0
50
100
150
200
250

**(c)**

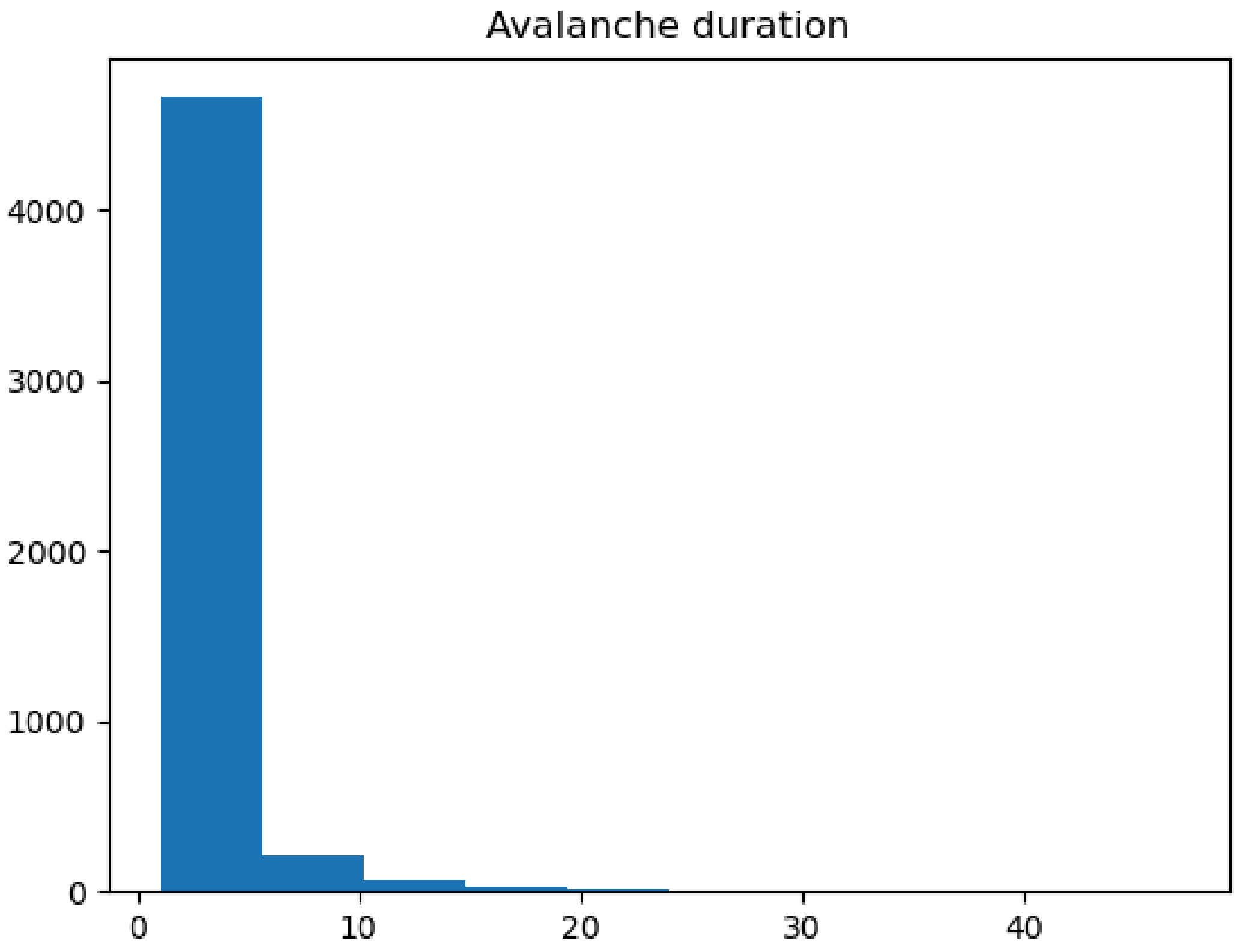

**(d)**

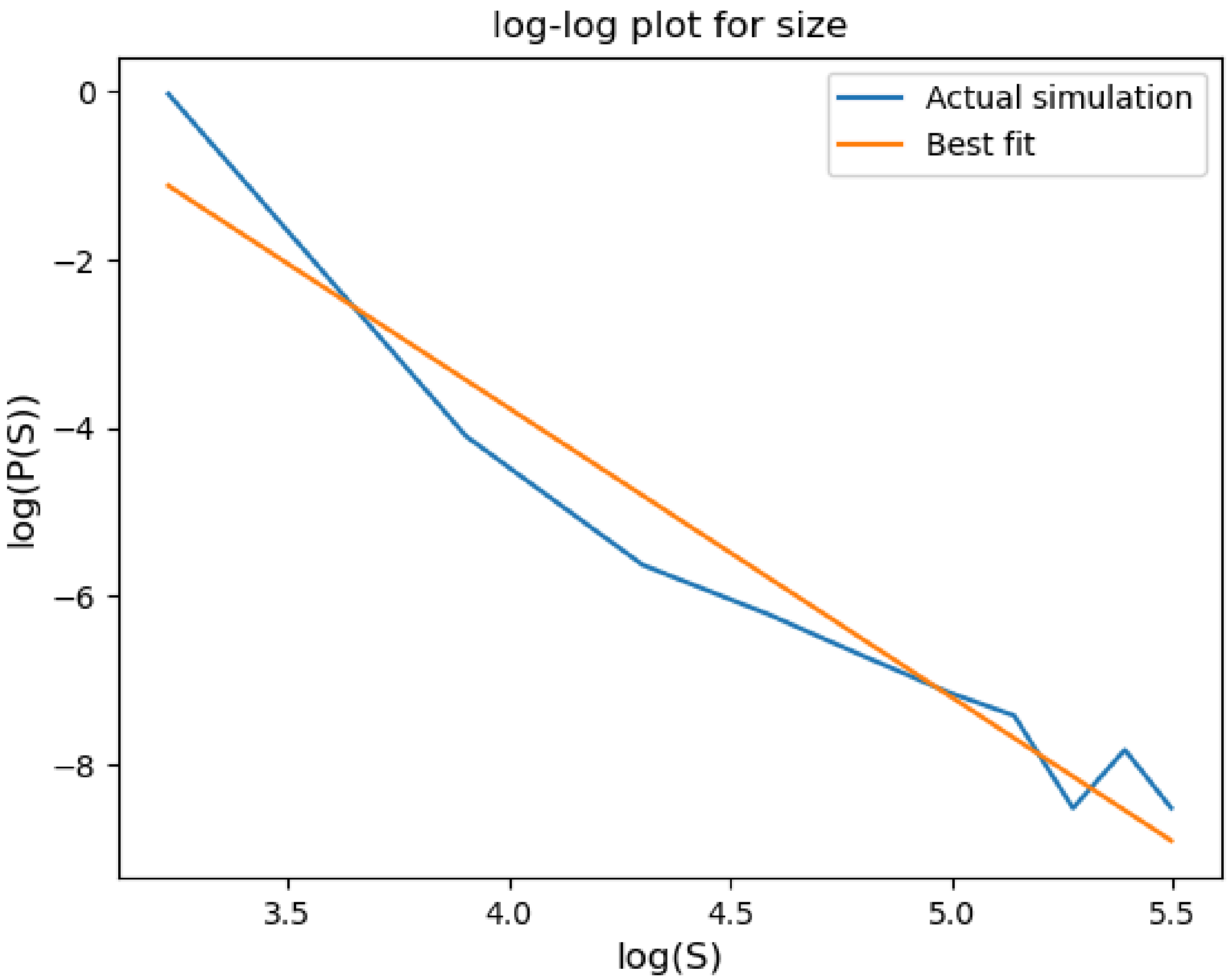
log-log plot for size
Actual simulation
Best fit
0
−2
−4
−6
−8
log(P(S))
3.5
4.0
4.5
5.0
5.5
log(S)

**(e)**

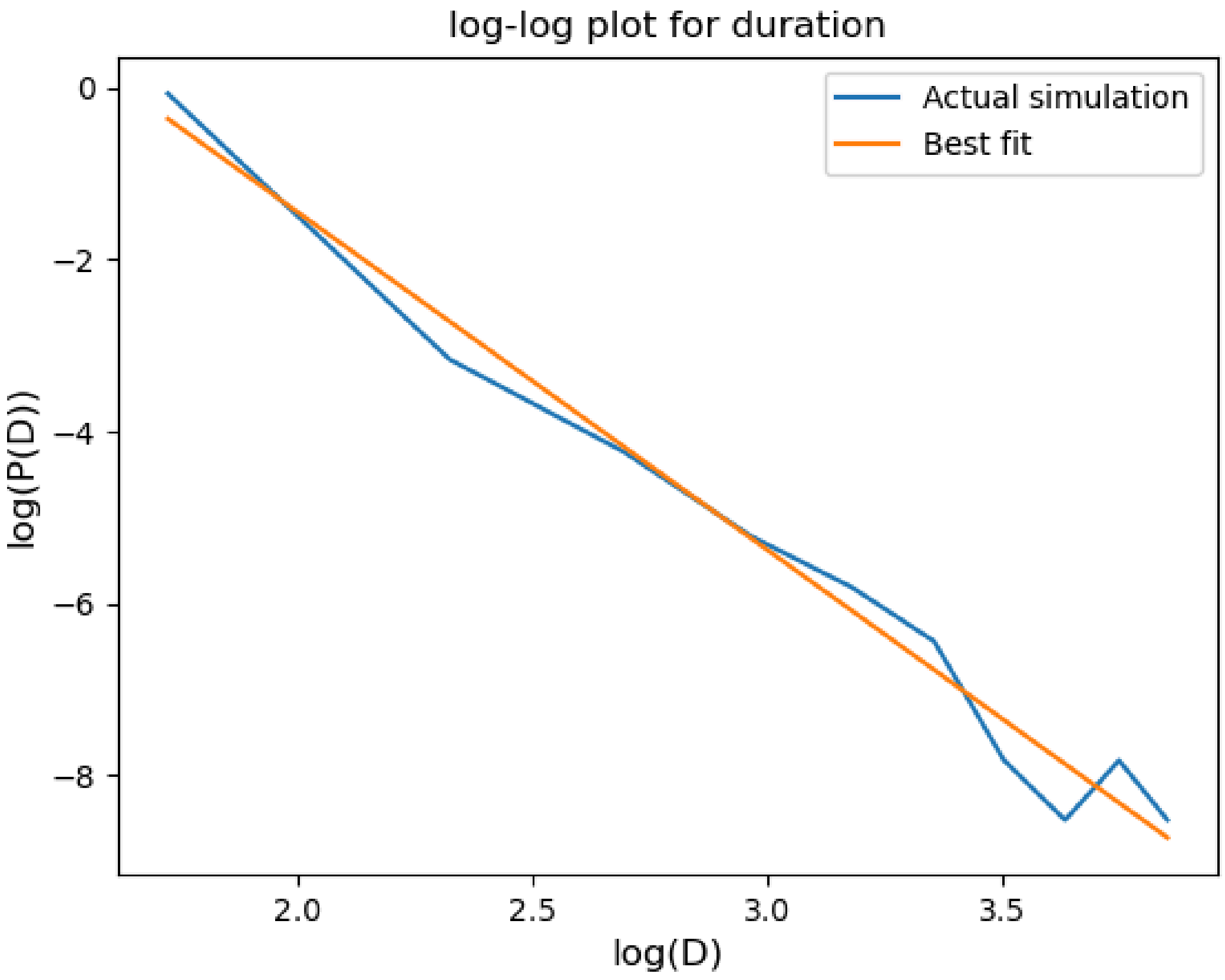
log-log plot for duration
Actual simulation
Best fit
log(P(D))
log(D)
0
−2
−4
−6
−8
2.0
2.5
3.0
3.5

**(f)**

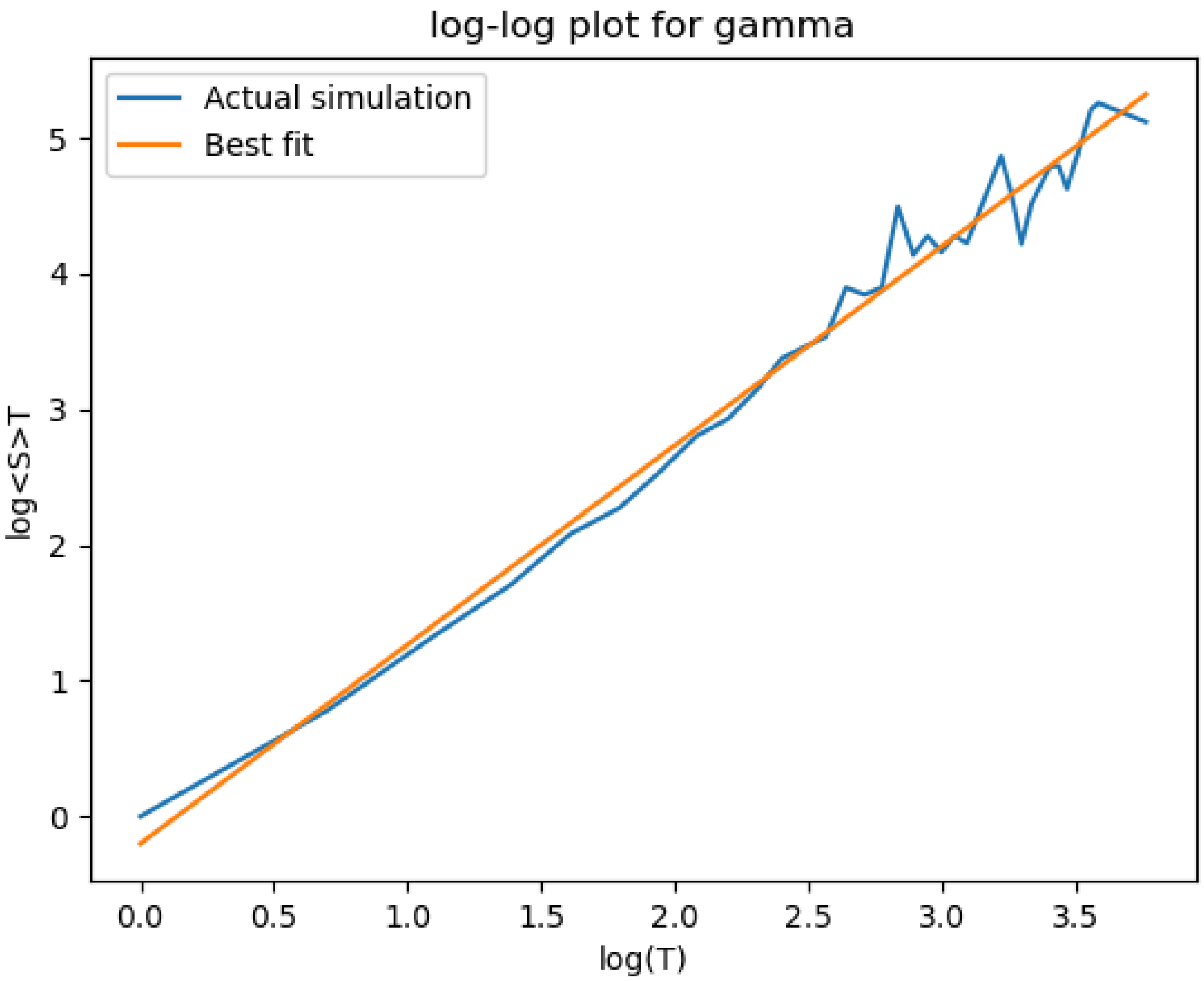
log-log plot for gamma
Actual simulation
Best fit
log<S>T
log(T)
0
1
2
3
4
5
0.0
0.5
1.0
1.5
2.0
2.5
3.0
3.5

**(g)**

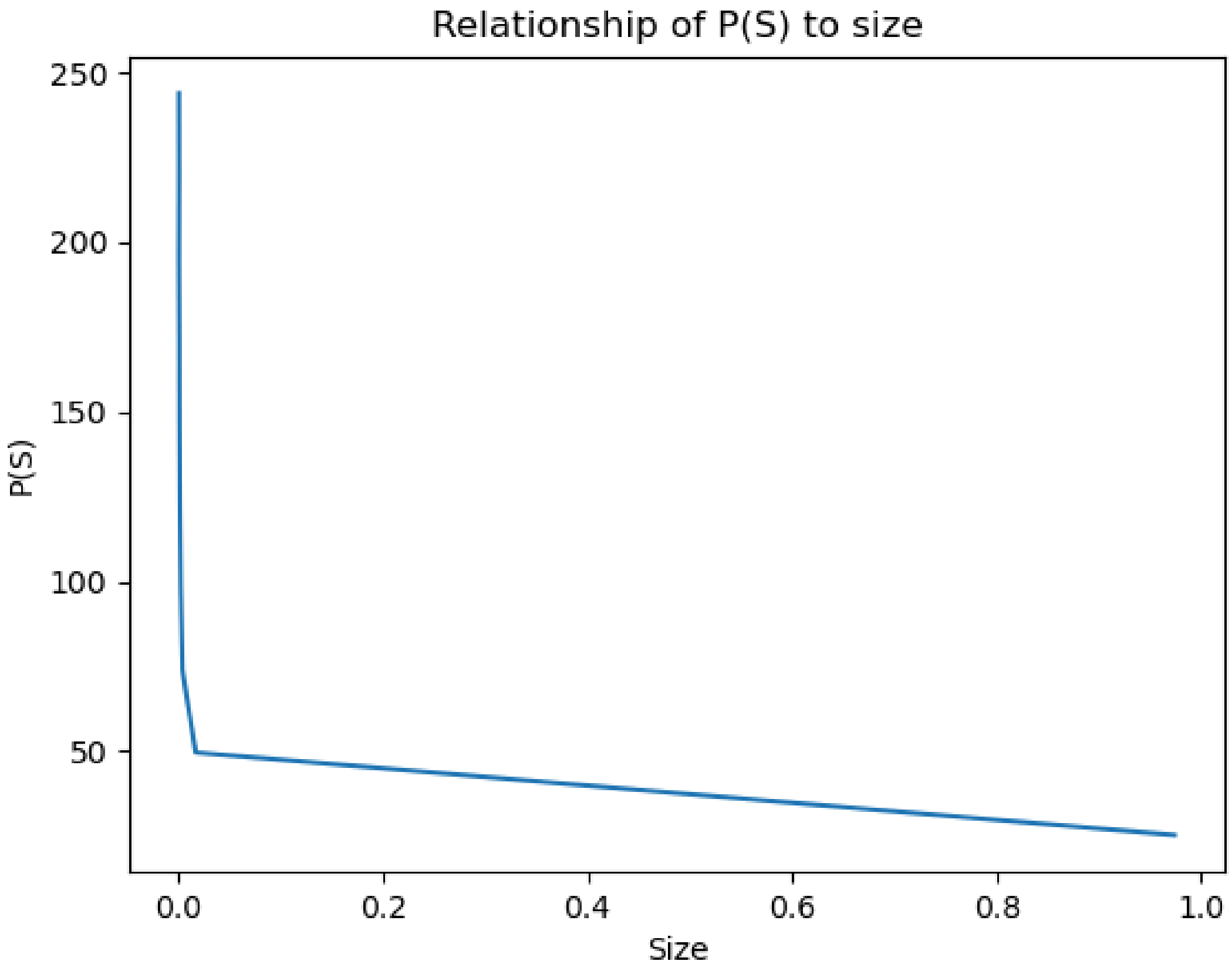
Relationship of P(S) to size
250
200
150
100
50
P(S)
0.0
0.2
0.4
0.6
0.8
1.0
Size

**(h)**

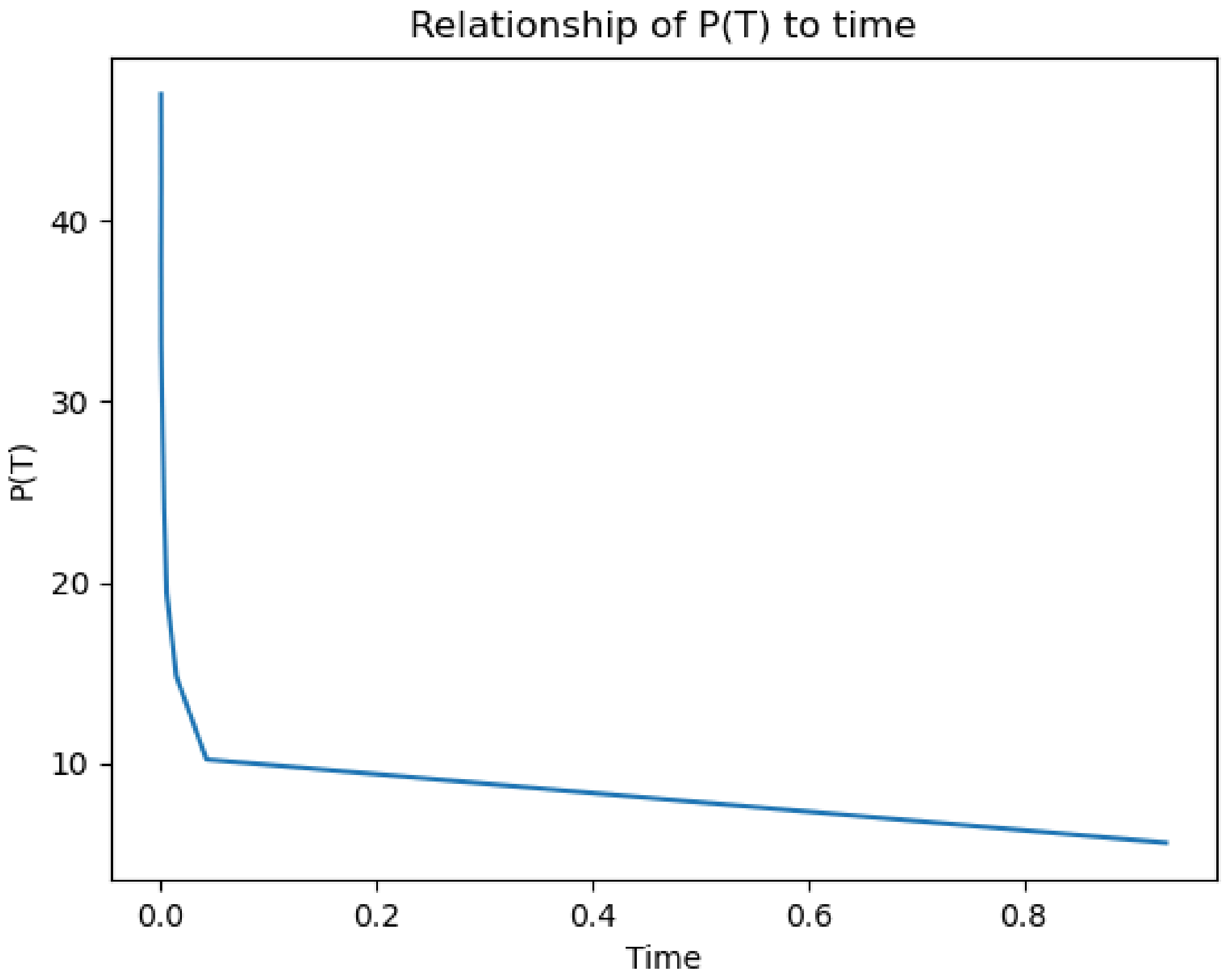

**(i)**

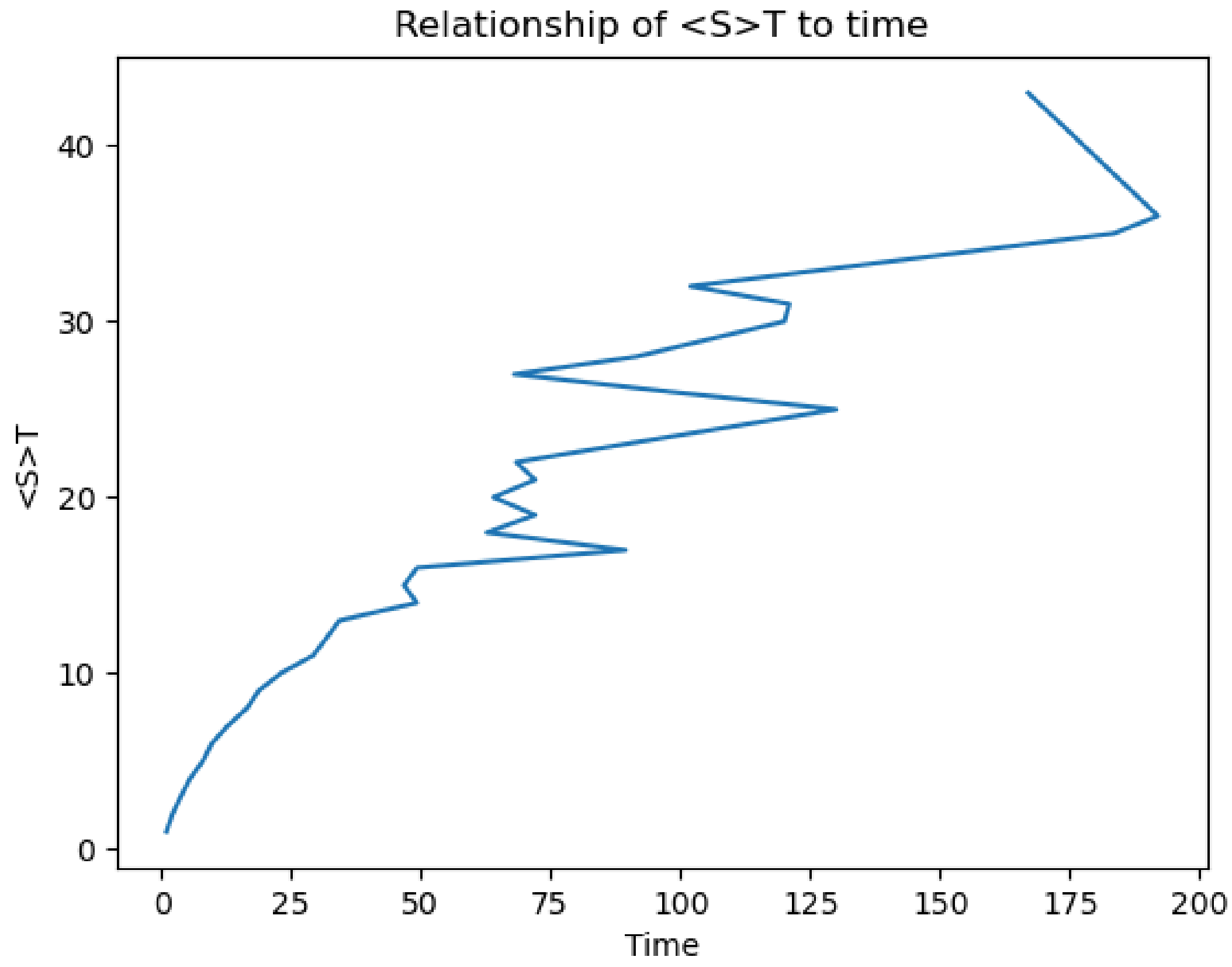

**(j)**

**Fig. 4 (a-j):** Sample branching process (branching ratio = 0.85)

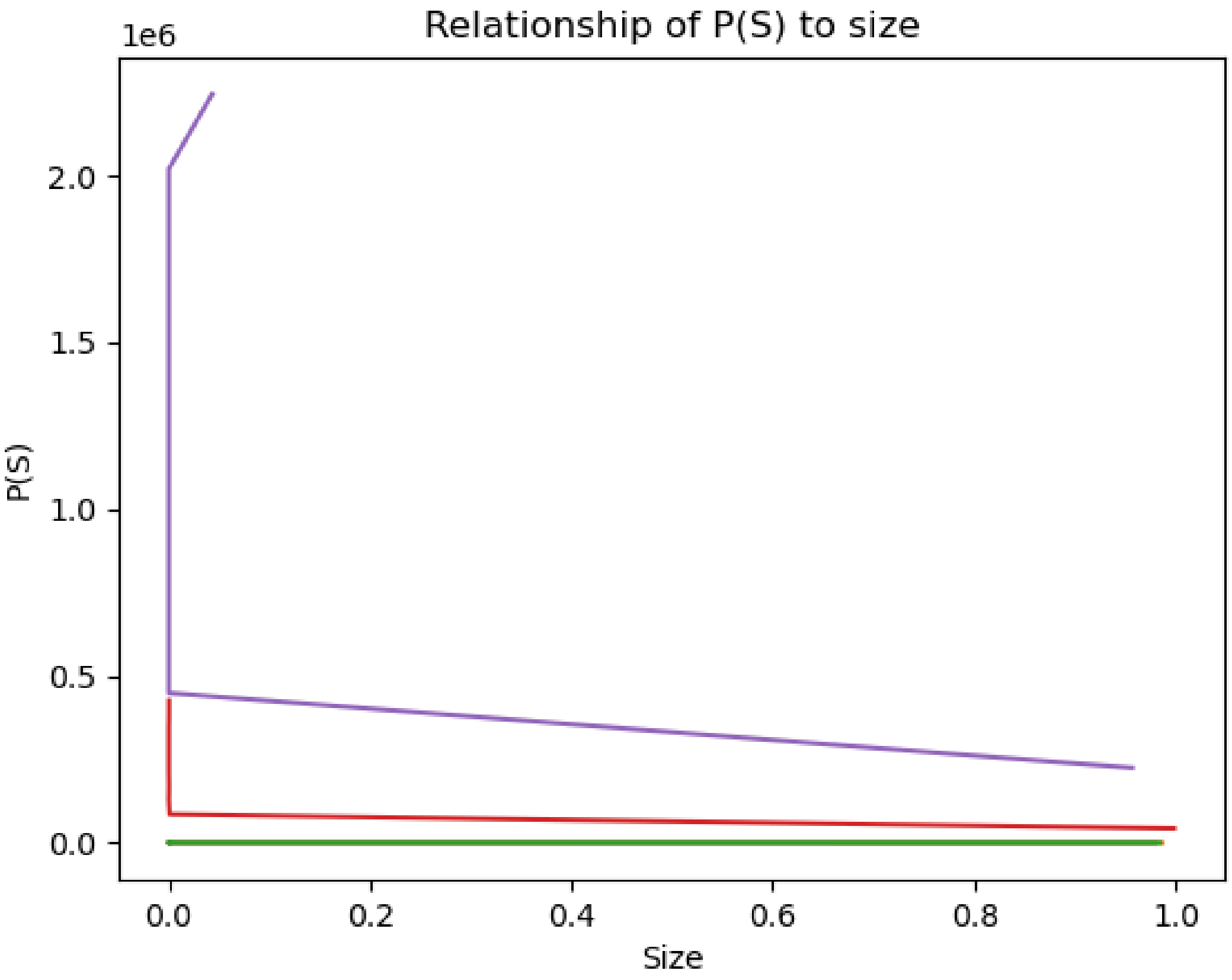

Relationship of P(S) to size
1e6
2.0
1.5
1.0
0.5
0.0
P(S)
0.0
0.2
0.4
0.6
0.8
1.0
Size

**(a)**

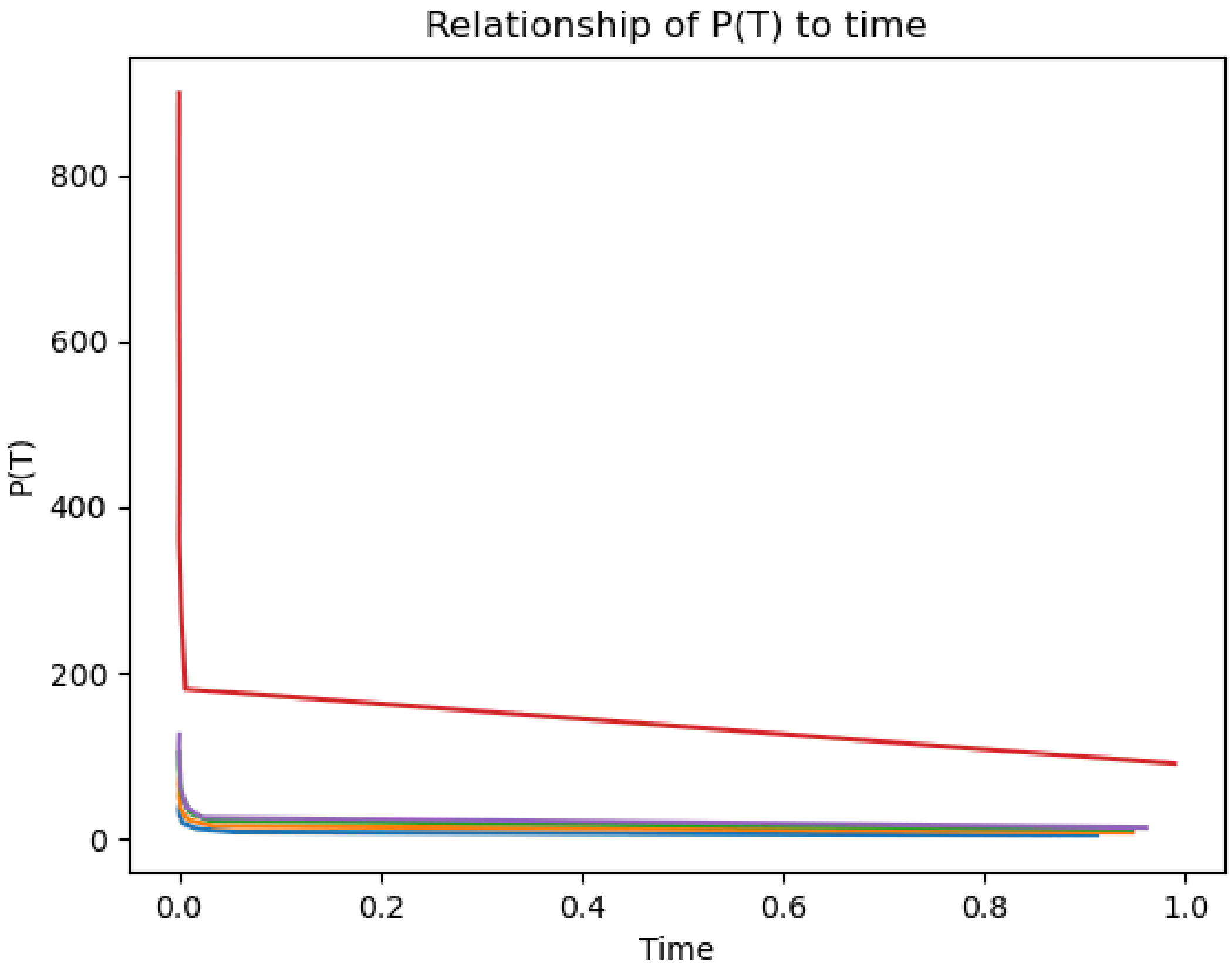
Relationship of P(T) to time
P(T)
800
600
400
200
0
0.0
0.2
0.4
0.6
0.8
1.0
Time

**(b)**

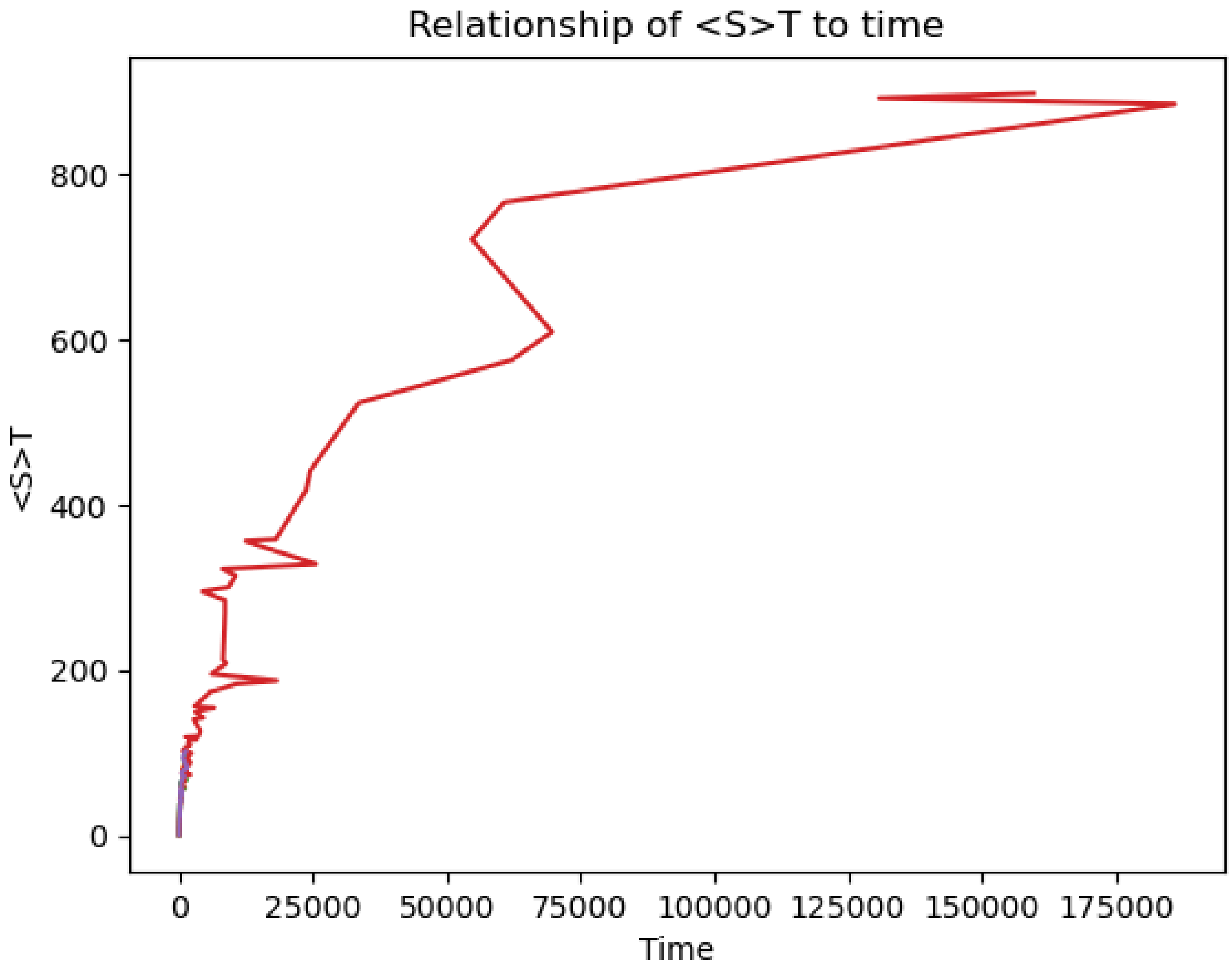

**(c)**

**Fig. 5 (a-c):** Branching processes with lambda from 0.85 to 1.05

There is experimental evidence for the log-normal distribution of neural responses (Buzsáki and Mizuseki, 2014; Koulakov, Hromadka and Zador, 2009) We wrote a function that was able to select responsive neurons, using a paired 't' test between firing rates before and after a stimulus event. The function worked better when a Bonferroni correction was applied (α/m) and also when the alpha term was divided by the square root of the number of comparisons performed (α/√m). This suggests that aggregates of neurons are the functional units of information storage, and that there is an inherent limit of resolution to our attempts to model it. The "decision" to form an avalanche is taken both at the level of the single neuron as well as the population. So, when we use coarse-graining methods to zoom in or zoom out, there is an inherent component of uncertainty or noise. One way to study this problem is through branching process simulations. As these methods approximate a very noisy process, this may show that they belong to the same universality class. The limit of resolution that is imposed on the system is like the measurement problem in quantum mechanics. Then, measurement is the unit process of a bit being defined in either of the two states +1 or -1 in unit time.

## Discussion

We trained 6 mice in a modified version of the 2-AFC task described by Burgess et al. (2017). We then recorded from the secondary motor cortices of the trained mice while they performed the task, using neuropixels probes. Then, we analyzed the behavioral and electrophysiological data to build a picture of the role of M2 in multisensory decision making.

We chose to study multisensory integration in mice as they are amenable to a large variety of experimental manipulations. The availability of new, powerful tools such as the Allen Brain Atlas and the Mouse Brain Architecture Project have provided maps of gene expression and brain connectivity that are not available for other species (Carandini and Churchland, 2013). Studies by Raposo et al. (2012) and Sheppard, Raposo and Churchland (2013) have shown that rodents combine multisensory information in a statistically optimal manner, like humans. Brunton, Botvinick and Brody (2013) have shown that rodents can accumulate information over time to make decisions based on an abstract quantity.

We chose to focus on the M2 as previous unpublished work in our lab used optogenetics to isolate the M2 as being involved in multisensory integration. While no such studies have been conducted involving M2, there have been some which have reported a role for other areas of the frontal cortex. Uylings, Groenewegen and Kolb (2003) and Kesner and Churchwell (2011) have shown that frontal cortex is necessary for higher cognitive functions in rodents. Other studies have shown that sub-areas such as the orbitofrontal cortex (Kepecs et al., 2008), agranular cortex (Erlich, Bialek and Brody, 2011) and anterior cingulate cortex (Narayanan and Laubach, 2006) are all involved in various aspects of motor planning and perceptual decision making. Studies by Angelaki, Gu and Deangelis, (2009) and Raposo, Kaufman and Churchland (2014) have reported a role for the posterior parietal cortex (PPC). However, in our optogenetic inactivation experiments, we found no role for the PPC in multisensory integration. Moreover, Raposo, Kaufman and Churchland (2014) themselves found that inactivating the PPC did not impair behavior in multisensory trials.

We chose a 2-AFC task design as it offers key advantages over other methods. Go/no-go tasks are a very popular study design but suffer from a key issue that they are highly vulnerable to changes in the animal's motivation and decision criterion. A decline in performance due to decrease in motivation could be misinterpreted as a failure to identify the relevant stimulus attribute (Carandini and Churchland, 2013). The 2-AFC design solved both these problems as in every choice, subjects must decide regarding the presence or absence of a stimulus attribute. Any bias towards one choice is readily detectable by analyzing the data.

The neuropixels probes themselves offer tremendous advantages over previously available methods of high-density electrophysiology (Jun et al., 2017). They have densely packed recording sites on a small 9 mm shank, which allows for high- spatiotemporal resolution with well isolated spiking units, along with a large volume coverage of brain regions (Harris et al., 2017). These high site count, single shank probes are a massive improvement over existing multi-shank silicon probes (Shobe et al., 2015; Rios et al., 2016; Scholvin et al., 2016). This allows for a small cross-sectional area which minimizes brain tissue damage.

On analyzing the data, we found neurons sensitive to auditory and visual stimuli as well as to the mode of stimulus presentation (active vs. passive). For all mice, neurons were more responsive to auditory than visual stimuli. Also, more neurons were responsive to either stimuli during active recordings, as compared to passive (3161 vs. 2195). This could be because animals are more engaged in the task, with higher motivation for reward, during the active mode. In the active recordings, all trials included an auditory stimulus, in the form of an onset tone. Perhaps, the mice are more sensitive to this stimulus, thus explaining the above effect. This could also be because the spiking data is contaminated with neuronal activity resulting from movement in the active mode. I tried to demarcate its effect on spiking, by using markers for movement onset in the cell raster. The example neurons shown clearly do not respond to movement, but the possibility does remain for the larger set of neurons.

We found a very low correlation between the neurons responding to auditory vs. visual stimuli, suggesting that specificity to one modality is independent of the other. Also, the fraction of cells differentiating between left and right sided stimuli was low in two of the three mice. The one mouse (mouse 2) which had a relatively high percentage was also the best performer during the training and recordings. I generally observed a marked decline in auditory performance after the second surgery (craniotomy). This decline was the least with mouse 2. This suggests that overcoming this limitation of the drop-in auditory performance may improve this number.

In conclusion, our studies have started to shed light on the possible mechanisms of how the M2 is involved in audiovisual integration.

We are essentially studying information transmission as a branching process. With the development of technology for large-scale neural recordings, it has become increasingly important to develop methods and tools to analyze this data. Methods from statistical physics and complexity science are now being applied to study the brain. Here, we have investigated these results further using simulations of neural activity as a branching process and tried to relate it to known properties of scaling relations.

Speculating further, we would want to study how these ideas relate to Alfred North Whitehead’s process philosophy (Lowe, 2020; Sherburne, 1966; Whitehead, 2010). According to Alfred North Whitehead’s process philosophy, processes hierarchically nested within processes are the fundamental building blocks of our world, and complexity science tries to further this viewpoint.

The general approach of complexity science can be defined as trying to study the processes between the components of a system, as opposed to processes inside components (Jensen, 2022). This approach may lead us to identify aggregate properties observed at multiple organizational levels.

**Limitations of methods**

One very prominent limitation in our study was the marked decline in performance in auditory trials following the craniotomy. This probably did affect the spiking activity of the neurons and prevents us from getting a complete picture of any possible correlations in the responses. We tried to overcome this by removing the ear-bars used during the surgery. Following this, the auditory performance did improve slightly, but the results were still not satisfactory. So, this is something that will require more efforts to solve.

Another limitation was the rapid re-growth of bone and scar tissue in and around the craniotomy. This greatly restricted the time available to us to record from the mice, before another craniotomy was needed, or the mouse had to be culled. We have tried various methods in the lab, such as placing a honeycomb disc around the borders of the craniotomy, to try and slow the growth, but none have been effective so far.

## References

1. Adrian, E. D., & Zotterman, Y. (1926). The impulses produced by sensory nerve endings: Part 3. Impulses set up by Touch and Pressure. The Journal of Physiology, 61(4), 465. https://doi.org/10.1113/JPHYSIOL.1926.SP002308
2. Angelaki, D. E., Gu, Y. and DeAngelis, G. C. (2009) 'Multisensory integration: psychophysics, neurophysiology, and computation', Current Opinion in Neurobiology, 19(4), pp. 452–458. doi: 10.1016/j.conb.2009.06.008.
3. Beggs, J. M. (2022). The Cortex and the Critical Point: Understanding the Power of Emergence. MIT Press.
4. Beggs, J. M., & Plenz, D. (2003). Neuronal Avalanches in Neocortical Circuits. Journal of Neuroscience, 23(35), 11167–11177. https://doi.org/10.1523/JNEUROSCI.23-35-11167.2003
5. Borst, A., & Theunissen, F. E. (1999). Information theory and neural coding. Nature Neuroscience 1999 2:11, 2(11), 947–957. https://doi.org/10.1038/14731
6. Brunton, B. W., Botvinick, M. M. and Brody, C. D. (2013) 'Rats and Humans Can Optimally Accumulate Evidence for Decision-Making', Science, 340(6128), pp. 95–98. doi: 10.1126/science.1233912.
7. Burgess, C. P. et al. (2017) 'High-Yield Methods for Accurate Two-Alternative Visual Psychophysics in Head-Fixed Mice', Cell Reports, 20(10), pp. 2513–2524. doi: 10.1016/j.celrep.2017.08.047.
8. Buzsáki, G., & Mizuseki, K. (2014). The log-dynamic brain: how skewed distributions affect network operations. Nature reviews neuroscience, 15(4), 264-278.
9. Calvert, G. A. and Thesen, T. (2004) 'Multisensory integration: methodological approaches and emerging principles in the human brain', Journal of Physiology- Paris, 98(1–3), pp. 191–205. doi: 10.1016/j.jphysparis.2004.03.018.
10. Carandini, M. and Churchland, A. K. (2013) 'Probing perceptual decisions in rodents', Nature Neuroscience, 16(7), pp. 824–831. doi: 10.1038/nn.3410.
11. Christensen, K., & Moloney, N. R. (2005). Complexity and criticality (Vol. 1). World Scientific Publishing Company.
12. Demas, J., Manley, J., Tejera, F., Barber, K., Kim, H., Traub, F. M., Chen, B., & Vaziri, A. (2021). High-speed, cortex-wide volumetric recording of neuroactivity at cellular resolution using light beads microscopy. Nature Methods 2021 18:9, 18(9), 1103–1111. https://doi.org/10.1038/s41592-021-01239-8
13. Erlich, J. C., Bialek, M. and Brody, C. D. (2011) 'A cortical substrate for memory-guided orienting in the rat', Neuron, 72(2), pp. 330–343. doi: 10.1016/j.neuron.2011.07.010.
14. Foss-Feig, J. H. et al. (2010) 'An extended multisensory temporal binding window in autism spectrum disorders', Experimental Brain Research, 203(2), pp. 381–389. doi: 10.1007/s00221-010-2240-4.

15. Fosque, L. J., Alipour, A., Zare, M., Williams-García, R. v, Beggs, J. M., Ortiz, G., & Ortiz, G. (2022). Quasicriticality explains variability of human neural dynamics across life span. https://arxiv.org/abs/2209.02592v1
16. Gordon, A., Banerjee, A., Koch-Janusz, M., & Ringel, Z. (2021). Relevance in the Renormalization Group and in Information Theory. Physical Review Letters, 126(24), 240601. https://doi.org/10.1103/PHYSREVLETT.126.240601/FIGURES/3/MEDIUM
17. Hahn, G., Petermann, T., Havenith, M. N., Yu, S., Singer, W., Plenz, D., & Nikolić, D. (2010). Neuronal avalanches in spontaneous activity in vivo. Journal of Neurophysiology, 104(6), 3312–3322. https://doi.org/10.1152/JN.00953.2009/ASSET/IMAGES/LARGE/Z9K0091003280008.JPEG
18. Harris, K. D. et al. (2017) ‘Accuracy of Tetrode Spike Separation as Determined by Simultaneous Intracellular and Extracellular Measurements’, Journal of Neurophysiology, 84(1), pp. 401–414. doi: 10.1152/jn.2000.84.1.401.
19. Iso, S., Shiba, S., & Yokoo, S. (2018). Scale-invariant feature extraction of neural network and renormalization group flow. Physical Review E, 97(5), 053304. https://doi.org/10.1103/PHYSREVE.97.053304/FIGURES/18/MEDIUM
20. Jensen, H. J. (1998). Self-organized criticality: emergent complex behavior in physical and biological systems (Vol. 10). Cambridge university press.
21. Jensen, H. J. (2022). Complexity Science: The Study of Emergence. Cambridge University Press.
22. Jun, J. J. et al. (2017) ‘Fully integrated silicon probes for high-density recording of neural activity’, Nature, 551(7679), pp. 232–236. doi: 10.1038/nature24636.
23. Kepecs, A. et al. (2008) ‘Neural correlates, computation and behavioural impact of decision confidence’, Nature, 455(7210), pp. 227–231. doi: 10.1038/nature07200.
24. Kesner, R. P. and Churchwell, J. C. (2011) ‘An analysis of rat prefrontal cortex in mediating executive function’, Neurobiology of Learning and Memory, 96(3), pp. 417–431. doi: 10.1016/j.nlm.2011.07.002.
25. Koulakov, A. A., Hromádka, T., & Zador, A. M. (2009). Correlated connectivity and the distribution of firing rates in the neocortex. The Journal of neuroscience, 29(12), 3685-3694.
26. Lowe, V. (2020). Understanding Whitehead. JHU Press.
27. Meshulam, L., Gauthier, J. L., Brody, C. D., Tank, D. W., & Bialek, W. (2019). Coarse graining, fixed points, and scaling in a large population of neurons. Physical Review Letters, 123(17), 178103. https://doi.org/10.1103/PHYSREVLETT.123.178103/FIGURES/5/MEDIUM
28. Murray, M. and Wallace, M. (2011) The Neural Bases of Multisensory Processes. CRC Press (Frontiers in Neuroscience). doi: 10.1201/b11092.

29. Narayanan, N. S. and Laubach, M. (2006) ‘Top-Down Control of Motor Cortex Ensembles by Dorsomedial Prefrontal Cortex’, Neuron, 52(5), pp. 921–931. doi: 10.1016/j.neuron.2006.10.021.
30. Ngampruetikorn, V., Bialek, W., & Schwab, D. (2020). Information-bottleneck renormalization group for self-supervised representation learning. Bulletin of the American Physical Society, Volume 65, Number 1.
31. Pachitariu, M. et al. (2016) ‘Fast and accurate spike sorting of high-channel count probes with KiloSort’, Advances in Neural Information Processing Systems, 2016.
32. Paradisi, P., Allegrini, P., Gemignani, A., Laurino, M., Menicucci, D., & Piarulli, A. (2013). Scaling and intermittency of brain events as a manifestation of consciousness. AIP Conference Proceedings, 1510(1), 151. https://doi.org/10.1063/1.4776519
33. Petermann, T., Thiagarajan, T. C., Lebedev, M. A., Nicolelis, M. A. L., Chialvo, D. R., & Plenz, D. (2009). Spontaneous cortical activity in awake monkeys composed of neuronal avalanches. Proceedings of the National Academy of Sciences of the United States of America, 106(37), 15921–15926. https://doi.org/10.1073/PNAS.0904089106
34. Quian Quiroga, R., & Panzeri, S. (2009). Extracting information from neuronal populations: information theory and decoding approaches. Nature Reviews Neuroscience 2009 10:3, 10(3), 173–185. https://doi.org/10.1038/nrn2578
35. Raposo, D. et al. (2012) ‘Multisensory Decision-Making in Rats and Humans’, Journal of Neuroscience, 32(11), pp. 3726–3735. doi: 10.1523/JNEUROSCI.4998-11.2012.
36. Raposo, D., Kaufman, M. T. and Churchland, A. K. (2014) ‘A category-free neural population supports evolving demands during decision-making’, Nature Neuroscience, 17(12), pp. 1784–1792. doi: 10.1038/nn.3865.
37. Rios, G. et al. (2016) ‘Nanofabricated Neural Probes for Dense 3-D Recordings of Brain Activity’, Nano Letters, 16(11), pp. 6857–6862. doi: 10.1021/acs.nanolett.6b02673.
38. Saxe, A. M., Bansal, Y., Dapello, J., Advani, M., Kolchinsky, A., Tracey, B. D., & Cox, D. D. (2019). On the information bottleneck theory of deep learning*. Journal of Statistical Mechanics: Theory and Experiment, 2019(12), 124020. https://doi.org/10.1088/1742-5468/AB3985
39. Schneidman, E., Slonim, N., Tishby, N., van Steveninck, R. deRuyter, & Bialek, W. (2001). Analyzing neural codes using the information bottleneck method. Advances in Neural Information Processing Systems, NIPS.
40. Scholvin, J. et al. (2016) ‘Close-packed silicon microelectrodes for scalable spatially oversampled neural recording’, IEEE Transactions on Biomedical Engineering, 63(1), pp. 120–130. doi: 10.1109/TBME.2015.2406113.
41. Schwartz-Ziv, R., & Tishby, N. (2017). Opening the Black Box of Deep Neural Networks via Information. https://doi.org/10.48550/arxiv.1703.00810
42. Seshadri, S., Klaus, A., Winkowski, D. E., Kanold, P. O., & Plenz, D. (2018). Altered avalanche dynamics in a developmental NMDAR hypofunction model of cognitive

impairment. Translational Psychiatry 2017 8:1, 8(1), 1–12. https://doi.org/10.1038/s41398-017-0060-z
43. Sheppard, J. P., Raposo, D. and Churchland, A. K. (2013) 'Dynamic weighting of multisensory stimuli shapes decision-making in rats and humans', Journal of Vision, 13(6), pp. 4–4. doi: 10.1167/13.6.4.
44. Shobe, J. L. et al. (2015) 'Brain activity mapping at multiple scales with silicon microprobes containing 1,024 electrodes', Journal of Neurophysiology, 114(3), pp. 2043–2052. doi: 10.1152/jn.00464.2015.
45. Sherburne, D. W. (1966). A key to Whitehead's process and reality.
46. Shew, W. L., Yang, H., Yu, S., Roy, R., & Plenz, D. (2011). Information Capacity and Transmission Are Maximized in Balanced Cortical Networks with Neuronal Avalanches. Journal of Neuroscience, 31(1), 55–63. https://doi.org/10.1523/JNEUROSCI.4637-10.2011
47. Song, W., Wang, J., Satoh, K., & Fan, W. (2006). Three types of power-law distribution of forest fires in Japan. Ecological Modelling, 196(3–4), 527–532. https://doi.org/10.1016/J.ECOLMODEL.2006.02.033
48. Song, Y.-H. et al. (2017) 'A Neural Circuit for Auditory Dominance over Visual Perception', Neuron, 93(4), pp. 940-954.e6. doi: 10.1016/j.neuron.2017.01.006.
49. Stein, R. B., Gossen, E. R., & Jones, K. E. (2005). Neuronal variability: noise or part of the signal? Nature Reviews Neuroscience 2005 6:5, 6(5), 389–397. https://doi.org/10.1038/nrn1668
50. Sterling, P., & Laughlin, S. (2015). Principles of neural design. MIT press.
51. Stevenson, R. A. et al. (2014) 'Identifying and Quantifying Multisensory Integration: A Tutorial Review', Brain Topography, pp. 707–730. doi:10.1007/s10548-014-0365-7.
52. Stringer, C., Michaelos, M., Tsyboulski, D., Lindo, S. E., & Pachitariu, M. (2021). High-precision coding in visual cortex. Cell, 184(10), 2767-2778.e15. https://doi.org/10.1016/J.CELL.2021.03.042
53. Tan, A., Meshulam, L., Bialek, W., Schwab, D., Tan, A., Meshulam, L., Bialek, W., & Schwab, D. (2019). The renormalization group and information bottleneck: a unified framework. APS, 2019, F66.007. https://ui.adsabs.harvard.edu/abs/2019APS..MARF66007T/abstract
54. Tishby, N., Pereira, F. C., & Bialek, W. (2000). The information bottleneck method. arXiv preprint physics/0004057.
55. Tishby, N., & Zaslavsky, N. (2015). Deep learning and the information bottleneck principle. 2015 IEEE Information Theory Workshop, ITW 2015. https://doi.org/10.1109/ITW.2015.7133169
56. Tagliazucchi, E., Balenzuela, P., Fraiman, D., & Chialvo, D. R. (2012). Criticality in large-scale brain fmri dynamics unveiled by a novel point process analysis. Frontiers in Physiology, 3 FEB, 15. https://doi.org/10.3389/FPHYS.2012.00015/BIBTEX

57. Uylings, H. B. M., Groenewegen, H. J. and Kolb, B. (2003) ‘Do rats have a prefrontal cortex?’, Behavioural Brain Research, pp. 3–17. doi: 10.1016/j.bbr.2003.09.028.
58. Whitehead, A. N. (2010). Process and reality. Simon and Schuster.
59. Williams-García, R. v., Moore, M., Beggs, J. M., & Ortiz, G. (2014). Quasicritical brain dynamics on a nonequilibrium Widom line. Physical Review E - Statistical, Nonlinear, and Soft Matter Physics, 90(6), 062714. https://doi.org/10.1103/PHYSREVE.90.062714/FIGURES/7/MEDIUM